\documentclass[aip,jcp,amsmath]{revtex4-2}
\usepackage[version=3]{mhchem} 
\usepackage{graphicx}
\usepackage{tabularx}
\usepackage{float}
\usepackage{amsmath}
\usepackage{amssymb}
\usepackage{easybmat}
\usepackage{booktabs}
\usepackage{siunitx}
\usepackage{multirow}
\usepackage{algorithm}
\usepackage{algpseudocode}
\usepackage{tikz}

\usetikzlibrary{positioning}
\usetikzlibrary{shapes,arrows,arrows,positioning,fit}
\usepackage{xcolor}
\usepackage{soul}
\usepackage[normalem]{ulem}
\renewcommand{\mathbf}[1]{\mbox{\boldmath$#1$\unboldmath}}

\begin{document}

\title{Full Quantum dynamics study for H atom scattering from graphene}

\author{Lei Shi}
\affiliation{Université Paris-Saclay, CNRS, Institut des Sciences Moléculaires d’Orsay UMR 8214, 91405, Orsay, France.}

\author{Markus Schröder}
\affiliation{Theoretische Chemie, Physikalisch-Chemisches Institut, Universität Heidelberg, Im Neuenheimer Feld 229, 69120, Heidelberg, Germany.}
\author{Hans-Dieter Meyer}
\affiliation{Theoretische Chemie, Physikalisch-Chemisches Institut, Universität Heidelberg, Im Neuenheimer Feld 229, 69120, Heidelberg, Germany.}
\author{Daniel Peláez}
\affiliation{Université Paris-Saclay, CNRS, Institut des Sciences Moléculaires d’Orsay UMR 8214, 91405, Orsay, France.}
\author{Alec M. Wodtke}
\altaffiliation{Institute for Physical Chemistry, Georg-August University of Göttingen, Tammannstraße 6, 37077 Göttingen, Germany.}
\affiliation{Dept. of Dynamics at Surfaces, Max Planck Institute for Multidisciplinary Sciences, am Faßberg 11, 37077 Göttingen, Germany.}
\author{Kai Golibrzuch}
\affiliation{Dept. of Dynamics at Surfaces, Max Planck Institute for Multidisciplinary Sciences, am Faßberg 11, 37077 Göttingen, Germany.}
\author{Anna-Maria Schönemann} 
\altaffiliation{Institute for Physical Chemistry, Georg-August University of Göttingen, Tammannstraße 6, 37077 Göttingen, Germany.}
\affiliation{Dept. of Dynamics at Surfaces, Max Planck Institute for Multidisciplinary Sciences, am Faßberg 11, 37077 Göttingen, Germany.}
\author{Alexander Kandratsenka}
\affiliation{Dept. of Dynamics at Surfaces, Max Planck Institute for Multidisciplinary Sciences, am Faßberg 11, 37077 Göttingen, Germany.}
\author{Fabien Gatti $^{\ast}$}%
 \email{fabien.gatti@universite-paris-saclay.fr} 
\affiliation{Université Paris-Saclay, CNRS, Institut des Sciences Moléculaires d’Orsay UMR 8214, 91405, Orsay, France.}
\date{\today}

\begin{abstract}
\textbf{\abstractname}\\
This study deals with the understanding of hydrogen atom scattering
from graphene, a process critical for exploring C-H bond 
formation and energy transfer during the atom surface collision. 
In our previous work (J.Chem.Phys \textbf{159}, 194102, (2023)), 
starting from a cell with 24 carbon atoms treated periodically, 
we have achieved quantum dynamics (QD) simulations with a
reduced-dimensional model (15D) and a simulation 
in full dimensionality (75D).  In the former work, the H atom attacked
the top of a single C atom, enabling a comparison of QD simulation
results with classical molecular dynamics (cMD). Our approach required
the use of sophisticated techniques such as Monte Carlo Canonical
Polyadic Decomposition (MCCPD) and Multilayer Multi-Configuration
Time-Dependent Hartree (ML-MCTDH), as well as a further development
of quantum flux calculations. We could benchmark our calculations by
comparison with cMD calculations. We have now refined our method to
better mimic experimental conditions. Specifically, rather than sending the H atom to a specific position on the surface, we have employed a plane wave for the H atom in directions parallel to the surface.  Key findings for these new simulations include the identification of discrepancies between classical molecular dynamics (cMD) simulations and experiments, which are attributed to both the potential energy surface (PES) and quantum effects. Additionally, the study sheds light on the role of classical collective normal modes during collisions, providing insights into energy transfer processes. The results validate the robustness of our simulation methodologies and highlight the importance of considering quantum mechanical effects in the study of hydrogen-graphene interactions.

\end{abstract}
\maketitle

\section{Introduction}
Graphene's exceptional electronic, magnetic, and chemical properties render it highly promising for a diverse range of applications \cite{Fang2020, Burnett1999, Novoselov2012, Schlapbach2001}. This has led to extensive research across scientific and technological domains, with chemical modification offering further avenues for innovation \cite{Lonkar2015}. For instance, hydrogenation of graphene, which is the most straightforward chemical modification, can create a local band gap, enabling the development of graphene-based semiconductor devices \cite{Balog2010}. Additionally, hydrogenated graphene serves as a valuable model system for investigating vibrational redistribution reactions \cite{Jiang2019}, and the formation of C-H chemical bond, presenting vast opportunities for exploration and discovery.

Hydrogen atom scattering from graphene surfaces is a pivotal area of research, holding significant implications for our understanding of fundamental processes like C-H bond formation and hydrogen storage mechanisms. This field of inquiry has been further propelled by recent experiments that have provided valuable insights into hydrogen adsorption on graphene, energy transfer during scattering events, and vibrational redistribution reactions \cite{Jiang2019}. In these groundbreaking experiments, hydrogen atoms were directed towards graphene surfaces at well defined energies and angles, and the kinetic energy of the scattered atoms was measured using advanced Rydberg atom time-of-flight (TOF) techniques.

\begin{figure}
    \centering
    \includegraphics[width=0.6\textwidth]{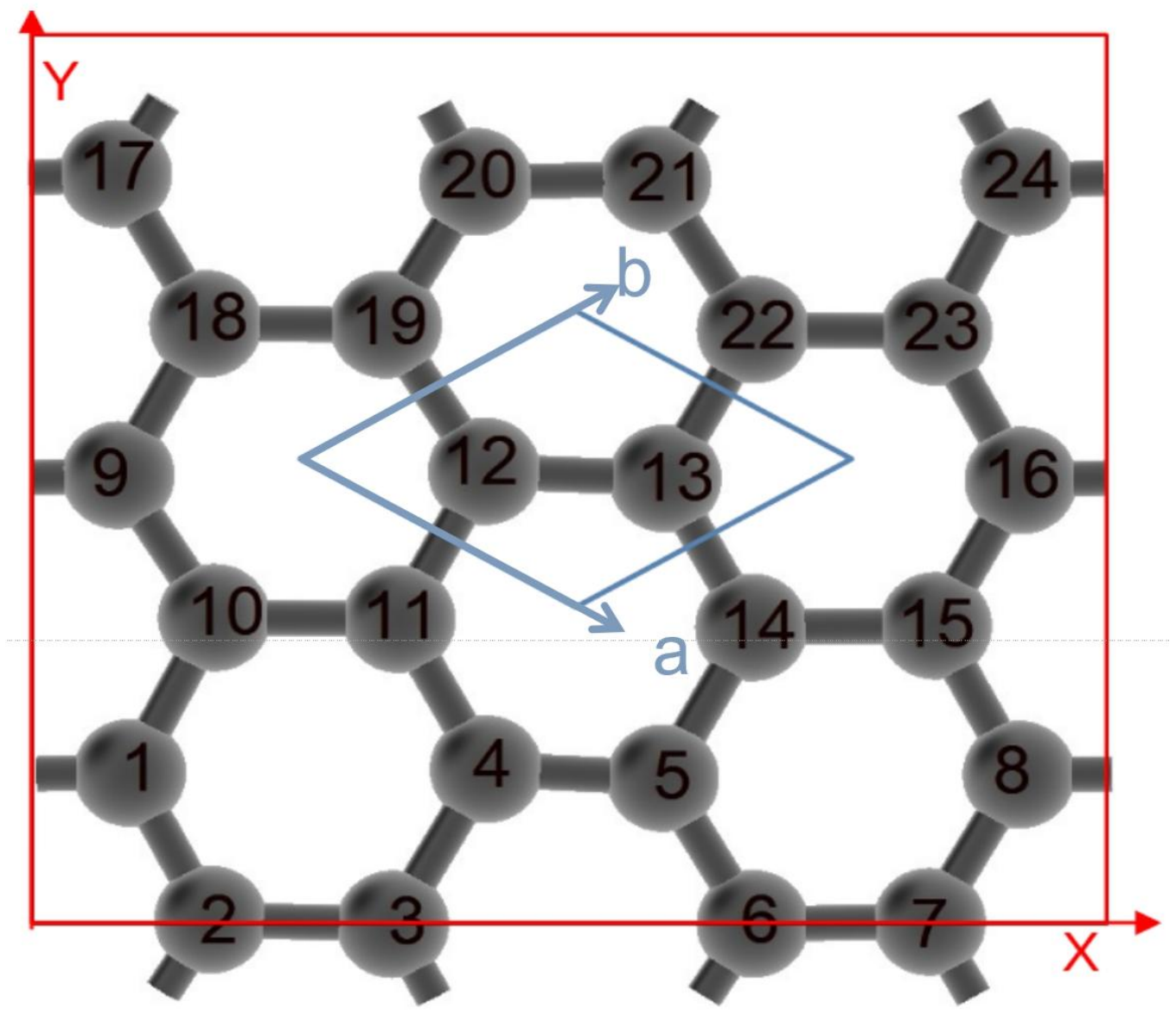}
    \caption{Graphene surface cell with the labels of the carbon atoms.}
    \label{fig:surf}
\end{figure}

To enhance our comprehension of these very complex experiments, a full-dimensional neural network potential energy surface (NN-PES) specifically tailored for the interaction between a hydrogen atom and graphene have been developed\cite{Wille2020}.  This NN-PES, constructed using state-of-the-art machine learning methods adapted for large condensed systems \cite{Behler2007}, has a periodic graphene cell comprising 24 carbon atoms, as depicted in Fig. \ref{fig:surf}, and a single free hydrogen atom. Molecular dynamics (MD) simulations have subsequently been performed utilizing this PES, yielding encouraging results that align well with experimental observations. 

Even though the MD simulations have demonstrated remarkable alignment with experimental findings under specific initial conditions, there are still some discrepancies stemming from the inherent quantum nature of the system depending on these initial conditions. treating such a complex and highly dimensional problem is very challenging.
In our previous study\cite{shi23:194102,shi23:059901}, we examined a reduced-dimensionality (15D) system, where only four of the 24 carbon atoms were allowed to move, they are carbon atoms 11, 12, 13, 19 as numbered in Fig. \ref{fig:surf}. The collisions between hydrogen atoms and graphene surfaces occurred only at the top of carbon atom 12. To accurately realize the quantum dynamics (QD) of this system, we employed sophisticated techniques such as the Monte-Carlo Canonical Polyadic Decomposition (MCCPD)\cite{mccpd} and the Multilayer Multi-Configuration Time-Dependent Hartree (ML-MCTDH) methods\cite{mey90:73,man92:3199,bec00:1,mey09:book} by using the \textit{Heidelberg MCTDH Package}\cite{mctdh:MLpackage}. We further established a benchmark for MD simulations based on the \textit{MD\underline{ }Tian2}\cite{mdt2git}, simulating the vibrational ground state, to ensure accurate comparisons.

Upon comparing the outcomes of our quantum dynamics simulations with the MD simulations, we observed a satisfactory agreement high in energy and for an incident hydrogen atom perpendicular to the surface, validating the methodologies used in the QD simulations. However, the slight deviations between these two simulation types highlight the existence of quantum effects. This underscores the importance of considering quantum mechanical effects when studying the intricate dynamics between hydrogen atoms and graphene surfaces: these differences may be very large for smallest energies and grazing hydrogen atoms.

In the first model of our previous article, the quantum effects were not very strong and the reduced dimensional system significantly differs from realistic experimental conditions. To address these limitations, we significantly enhanced our approach by increasing the dimensionality
of the system to 75D, enabling the movement of all C atoms shown in
Fig. \ref{fig:surf}. However, the H atom was still attacking the top
of a single C atom. In the present work, we have now refined our method even more to better mimic experimental conditions. Specifically, rather than sending the H atom to a specific position on the surface, we have employed a plane wave for the H atom in directions parallel to the surface. This approach, which replaces the Gaussian function mentioned in our previous article by a plane wave, allows for a more realistic simulation of the interaction between the H atom and the C atoms. However, this modification has also led to a significant increase in the complexity of our calculations due to the substantially greater correlation between the various degrees of freedom (DOFs) when compared to the previous simulation where the H atom only attacked a single carbon atom. This leads to a very correlated quantum system in 75 dimensions since, now, all the carbon atoms can be strongly coupled to the movement of the hydrogen atoms. There is no hierarchy any more between the carbon atoms. 
As a first step, in order to simplify the calculations, we have
thus made an additional approximation by choosing a smaller cell
for the H atom that is also periodic. Moreover, we still work with
an incident hydrogen atom perpendicular to the surface.

By adopting this improvement (using plane waves), we aim at achieving a more realistic representation of the dynamics of the system, ultimately enabling us to draw more reliable conclusions and insights from our simulations compared directly with the experimental results.

\section{\label{sec:methods}Classical MD and QD simulations details}

\subsection{Coordinates}

The cMD simulations were done using Cartesian coordinates. As in our previous article, in the QD simulations, the graphene surface are described by normal modes calculated by diagonalization of the Hessian matrix of the PES when the H atom is far from the surface. 72 normal modes are found for 24 carbon atoms, 3 of them have zero frequency due to the translation ensemble of the periodic surface. These 3 translation normal modes are removed from following dynamic
simulations. Hence, the propagation is actually in 72 dimensions, but
we keep the notation 75D in the following since the neglect of the
three translational modes does not establish an approximation.
More details are given in \textit{appendix} \ref{apx:a}.

\subsection{Initial conditions}
For the graphene surface, in the QD simulations, a negative imaginary time propagation\cite{mey06:179} is performed first to relax the graphene surface to its vibrational ground state in order to ensure that the collision propagation will begin from a well defined state. In other words, the graphene surface temperature is 0 $K$. 

We employed the same method introduced in our previous  article to generate 1,000 trajectories of the C atoms within a distribution which can mimic the vibration ground state wave function for cMD simulations\cite{shi23:194102}. The zero point energy calculated in the MCTDH is 4.05 $eV$, and the average energy for those trajectories used in the cMD simulations is 4.06 $eV$. Thus, we perform quasi-classical simulations instead of purely classical simulations.  As shown below, this sole change makes an important improvement of the classical simulations. 

As aforementioned, in order to make our simulations closer to the experimental conditions, we thus allowed the H atom to attack anywhere of the graphene surface, in stead of attacking only at the top of one C atom. Obviously, this makes the simulations much more realistic. 

{We still focus on the perpendicular collision, the incident angle is 0$^{\circ}$ in this paper.} In cMD simulations, we used 20,000 H atom trajectories, their initial X and Y position will be randomly generated according to a uniform distribution. In QD simulations, we used a plane wave for the H atom in the X and Y direction with periodic boundary conditions. As we are studying normal incidence scattering, the
initial X and Y momenta are zero and the plane wave turn into a
constant in X- and Y-direction. The H atom wave function in
Z-direction is a Gaussian function multiplied with an exponential
which defines the mean momentum. This leads to a distribution of
the initial kinetic energy of the H atom which is roughly of Gaussian
shape. However, in the following result analysis step, we make a flux
calculation in which we select an initial kinetic energy and analyze the wave function for this specific kinetic energy\cite{ris93:4503,ris96:1409,jae95:5605}. Thus, a precise definition of the shape of the Gaussian distribution (such as the width of the Gaussian function) is not so important in our study.  

\subsection{PES refitting}
We used MCCPD\cite{mccpd} to refit the original PES in order to transform it into the sum of products (SOP).


The PES $V(q_1,q_2,\ldots,q_f)$ is approximated by a canonical polyadic decomposition (CPD) form: 
\begin{equation}
    V^{\rm CPD}(q_1,q_2,\ldots,q_f) =\sum_{r=1}^Rc_r\prod_{\kappa=1}^{f}v_\kappa^r(q_\kappa) \,,
    \label{eq:cpd}    
\end{equation}
or in the grid-representation after discretization of the coordinates $\mathbf{q}$:
\begin{equation}
    V^{\rm CPD}_{i_1,i_2,\ldots,i_f} =\sum_{r=1}^Rc_r\prod_{\kappa=1}^{f}v^r_{\kappa,i_\kappa} \,,
    \label{eq:cpd_grid}
\end{equation}
where index $i_\kappa$ numbers discretization points of the coordinate $q_\kappa$, $v^r_{\kappa}$ are basis functions
with expansion coefficient $c_r$, and $R$ is the expansion order (the rank of the decomposition).  
It should be noted that the basis functions $v^r_{\kappa}$ are normalized but not orthogonal. 

Introducing the composite index $I=\{i_1,\ldots,i_f\}$, the decomposition is obtained by optimization of the
weighted $\mathcal{L}^2$ difference between the fitting potential $V_I^{\rm CPD}$ from Eq.~(\ref{eq:cpd_grid})
and the exact potential $V_I$. This can be cast into minimizing the functional
\begin{equation}
    J  = \sum_I W_I\left( V_I - V_I^{\rm CPD}\right)^2 + \epsilon  \sum_{I,r} W_I ~ c_r^2 
          \left(\prod_{\kappa=1}^f v^r_{\kappa,i_\kappa} \right)^2, 
    \label{eq:cpd_functional}
\end{equation}
where $W$ is a positive weight function depending on all coordinates $\mathbf{q}$, which imposes more weights on the regions the PES that should be fitted with elevated accuracy (for instance, low energy regions where the wavefunction resides). The second term in the r.h.s of Eq.\ (\ref{eq:cpd_functional}) proportional to $\epsilon$ serves as a regularization  with $\epsilon$ being a small parameter { which is set to the square root of machine precision, i.e., $\epsilon\approx 1\cdot 10^{-8}$}. It preserves the linear independence of the basis functions $v^r_\kappa$ during the optimization process. 

The functional $J$ is minimized by a variant of the \emph{alternating least squares} (ALS)\cite{com09:393,lat08:1067} algorithm.
The ALS iterations are computationally feasible only when the number of points $I$ to be processed is not larger than $10^{9}$--$10^{10}$. To tackle the system considered in this work, where the total number of points is of order of $10^{20}$,
we minimize the functional $J$ only on a set of Monte-Carlo sampling points with a distribution functions defined by the
weight $W$, i.e. the weight is implemented by a non-uniform distribution of sampling points\cite{mccpd}.

The sampling strategy, i.e. the choice of $W$, has proven to affect strongly the accuracy of the fit. For example,
Metropolis sampling has led to satisfying results\cite{mccpd,Schroeder2022}. Here, the distribution function is 
proportional to $\exp(-\beta V)$ with $\beta$ being a suitable positive parameter. It emphasizes low energy regions
and covers the area where the wavefunction mostly resides. However, the sampling distribution is derived from the
potential part of the total Hamiltonian only and neglects the influence of the kinetic energy operator. 
A much better choice is hence the weight
\begin{eqnarray}
 W (\mathbf{q}) &=& Z^{-1}\,\langle \mathbf{q}|\exp(-\beta H)| \mathbf{q}\rangle \nonumber \\
   &=& Z^{-1}\, \sum_n \exp(-\beta E_n) |\psi_n(\mathbf{q})|^2
   \label{eq:WZ}
\end{eqnarray}   
where $(E_n, \psi_n)$ denote the eigenpairs of the Hamiltonian $H$, and where $Z=tr(e^{-\beta H})=\sum_n \exp(-\beta E_n)$
is the partition function, inserted to keep the distribution normalized. This distribution, however, is difficult
to evaluate numerically. But when one lets $\beta$ go to infinity, one arrives at
\begin{equation}
   W (\mathbf{q}) = |\psi_0(\mathbf{q})|^2
   \label{eq:W0}
\end{equation}
i.e. the density of the ground state wavefunction. A set of sampling points representing this distribution is conveniently
generated by diffusion Monte-Carlo (DMC)\cite{Kosztin1996} simulations. We performed two DMC simulations, one to obtain
the vibrational ground state distribution of graphene not interacting with a projectile (H atom is far away from the 
surface) and another one for graphene with the H atom close to it. The sampling points obtained from both calculations
were  subsequently merged into a single set and each sampling point was mapped onto the nearest Discrete Variable
Representation (DVR) point. 

The ALS algorithm is a iterative procedure, one has to start with an initial guess for the CPD decomposition.
This decomposition is then improved by an ALS step and serves as input for the next iterartion.
A efficient way to accelerate convergence is to start with a small rank $R$, perform a number of iterations
and subsequently increase $R$, where the new coefficients $c_r$ are set to zero and the new basis
functions $v^r_\kappa$ are filled with random numbers. For details see Ref.[\!\!\citenum{mccpd}].

As plane waves are used initially to describe the $XY$ motion, the H atom wavefunction will explore a large 
area of the PES, thus more samplings have to be used for MCCPD calculation. For example, if the H atom wave
function in X and Y direction is restricted to the red rectangle region shown in Fig. \ref{fig:surf},
one needs about 24 times more samplings than our previous MCCPD in which we concentrated the H attack to only
one C atom. The increase of sampling points will increase the memory and CPU-time of refitting and makes
the convergence of MCCPD calculation very difficult. 

We decided to reduce the region for the H atom in the X and Y directions. We will allow the H atom be periodic in the 
blue rhombus region defined by the twisted coordinate $a$ and $b$. The blue rhombus region is an elementary cell of the
graphene, it contains all possible collision positions. The transformation of twisted
coordinates from the Cartesian coordinate of the H atom in the X and Y direction is given by
\begin{equation}
\begin{pmatrix}
x \\
y
\end{pmatrix}
=
\begin{pmatrix}
\cos(\pi/6) & \cos(\pi/6) \\
-\sin(\pi/6) & \sin(\pi/6)
\end{pmatrix}
\times
\begin{pmatrix}
a\\
b
\end{pmatrix} ,
\end{equation}
the first derivatives are:
\begin{align}
     \frac{\partial}{\partial x}&=\frac{1}{2\cos(\pi/6)}\frac{\partial}{\partial a}+\frac{1}{2\cos(\pi/6)}\frac{\partial}{\partial b}; \nonumber \\
     \frac{\partial}{\partial y}&=-\frac{1}{2sin(\pi/6)}\frac{\partial}{\partial a}+\frac{1}{2sin(\pi/6)}\frac{\partial}{\partial b},
\end{align}
and the new kinetic operator for the H atom in the plane after transformation is:
\begin{align}
    \hat{T}&=-\frac{1}{2m}\frac{\partial^2}{\partial x^2}-\frac{1}{2m}\frac{\partial^2}{\partial y^2} \nonumber \\
    &=-\frac{1}{2m}\frac{1}{4\cos(\frac{\pi}{6})^2\sin(\frac{\pi}{6})^2}\frac{\partial^2}{\partial a^2}-\frac{1}{2m}\frac{1}{4\cos(\frac{\pi}{6})^2\sin(\frac{\pi}{6})^2}\frac{\partial^2}{\partial b^2}\nonumber \\
    &-\frac{1}{2m}\frac{\sin(\frac{\pi}{6})^2-\cos(\frac{\pi}{6})^2}{2\cos(\frac{\pi}{6})^2\sin(\frac{\pi}{6})^2}\frac{\partial^2}{\partial a \partial b},
\end{align}
where $x$ and $y$ represent the original Cartesian coordinates, while $a$ 
and $b$ correspond to the new coordinates, and $m$ is the mass of the H atom.
By utilizing these new coordinates for the H atom in the XY plane, we can
focus solely on selecting samplings from the DMC calculation for the H atom
being within the blue elementary cell of Figure \ref{fig:surf}.

Restricting the parallel motion of the H-atom to the small $a$-$b$ unit cell is correct (the "blue" cell in Fig. \ref{fig:surf}), if the surface is rigid.
When the H-atom leaves the unit cell on one side, it enters it again on the opposite side, due to periodic boundary
conditions. Doing so it suffers from exactly the same potential, as if when it would have continued its movement
out of the unit cell. This is because the surface is periodic with respect to the unit cell. However, when the
H-atom interacts with the surface, the surface atoms start to move and the periodicity is no longer strictly obeyed.
Hence, our approach is an approximation since it enforces some constraints on the movements of the atoms outside the blue cell. To improve the approximation it is planned to double the lengths of $a$
and $b$, hence quadruple the area of the cell. This, however, is a major challenge, because with the present
system we are already close to the limit of what is possible with our present hardware. However, our model is, as aforementioned, a major improvement compared to the former one (targeting one carbon atom) since it allows the hydrogen atom to target all the geometries in the blue cell, in particular the geometries between the carbon atoms. 

\begin{figure}
     \centering
     \includegraphics{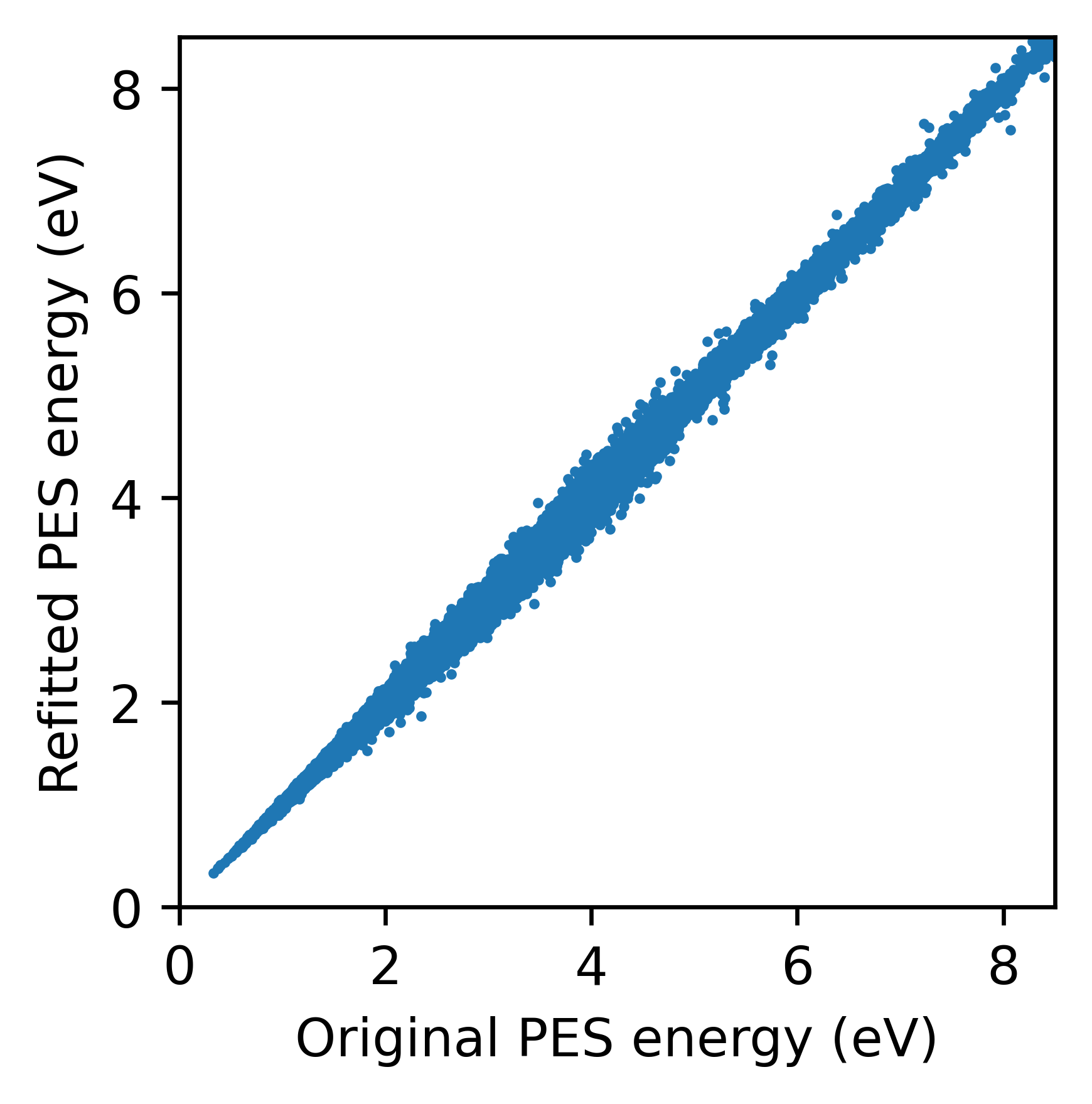}
     \caption{Orignial PES versus the refitted PES of full dimensional system.}
     \label{fig:cpvsorg}
 \end{figure}

 The original PES, which is in Cartesian coordinates, is transformed to the normal coordinates during the refitting. A total of 368,640 sampling points were generated for the above setting, involving 81,920 samplings calculated by DMC calculations targeting the hydrogen atom located on carbon atoms with the numbers 12 and 13 in Fig. \ref{fig:surf}, 122,880 samplings calculated by DMC calculation with the hydrogen atom positioned far from the graphene surface, 102,400 samplings chosen from a cMD simulation when the initial kinetic energy is 1 $eV$, and 61,440 samplings chosen from a cMD simulation when the initial kinetic energy is 2 $eV$. The optimization process initiated with a rank of 256 and incrementally increased by 256 after every 30 iterations until achieving a final rank of 1792. The refit is much better converged than before, the RMS error is 223 $cm^{-1}$ after optimization. It is already at chemical precision, i.e. smaller than 1 $kcal/mol$. Similarly, we tested the refitted PES with new samplings from new calculations with the same conditions and methods. The original PES versus the refitted PES is with the test samplings as shown in figure \ref{fig:cpvsorg}. Nevertheless, there is room for improvement in the refitting quality around the 4 $eV$ region. Increasing the number of samplings and elevating the rank could enhance refitting quality, but this leads to an increased computational cost.

\subsection{ML-MCTDH}

In the MCTDH method \cite{mey90:73,man92:3199,bec00:1,mey09:book}, the molecular wavefunction is 
formulated as the sum of product (SOP) of orthonormal time-dependent single particle functions (SPFs) of a set of logical coordinates $Q_1,Q_2,...,Q_p$:
\begin{equation}
    \Psi(Q_1,\ldots,Q_p,t)
    =\sum_{i_1=1}^{n_1}\cdots\sum_{i_p=1}^{n_p} A^1_{i_1,\ldots,i_p}(t)\,
    \varphi^{1;1}_{i_1}(Q_1,t)\cdots\varphi^{1;p}_{i_p}(Q_p,t),
    \label{eq:mctdh}
\end{equation}
with SPFs defined as a linear combination of the primitive time-independent basis, such as,
\begin{equation} 
\varphi^{1;\kappa}_m(Q_\kappa,t)=\sum_{j_1,...,j_d} A^{2;\kappa}_{m,j_1,...,j_d}
(t) \,\chi_{j_1}^{\kappa,1}(q_m^1)\cdots\chi_{j_d}^{\kappa,d}(q_m^d) \,.
    \label{eq:spf}
\end{equation}
Here $n_p$ is the number of SPFs used for the $p$th combined mode, $d$ degrees of freedom (DoFs) are combined to form the $m$th logical coordinate, $Q_m = (q_m^1,\ldots,q_m^d)$. The logical coordinates are able to improve the convergence of the calculation by treating the correlation between the combined DoFs already on the SPF level.

One should carefully chose the type and the number of the primitive basis functions $\chi_{j_d}^{\kappa,d}(q_m^d)$ in order to adapt to the nature of the system and to converge the calculations. Table \ref{tab:dvr75D} lists the primitive bases used in the 75D QD simulation. We used a Fast Fourier
Transform method (FFT) for the 3 coordinates of the H atom. This is
because the H atom in Z-direction can move in vast domain, the number
of grid points is large, whereas we should ensure periodic boundary
conditions in the directions parallel to the surface. For the normal
modes of graphene surface, we used the harmonic oscillator (HO) DVR. The frequency and mass used for HO basis function in Table \ref{tab:dvr75D} are equal to 1.0, because the normal modes used here are dimensionless, mass and frequency are considered in the propagation operator. It is important to note that the normal modes of graphene were calculated when the H atom is far from the surface as described in the \ref{apx:a}. However, the hydrogen atom can strongly alters the harmonic nature of the graphene structure when it approaches to the surface. Consequently, more primitive basis functions are required to achieve a converged calculation.

\begin{table}[t!]
\caption{The parameters for the definition of primitive basis functions. The first column is the name of DoFs, the following columns are the type of DVR, number of grid points, parameters of the DVR corresponds. In the case of FFT, they correspond to the coordinates of the first {(P1)} and last {(P2)} grid points and if it is periodic, it is written in the column (P3). In the case of HO, they correspond to the equilibrium position (P1), frequency (P2) and mass (P3) of the harmonic oscillator basis functions.}
    \centering
    \begin{tabularx}{\textwidth}{|>{\centering\arraybackslash}X|>{\centering\arraybackslash}X| >{\centering\arraybackslash}X| >{\centering\arraybackslash}X| >{\centering\arraybackslash}X| >{\centering\arraybackslash}X|}
    \hline
     DOF & Function type & Grids number & Parameter 1 & Parameter 2 & Parameter 3 \\
 \hline
  $H_a$ &    FFT   &    48   &   0.0  &   4.66286 &  periodic \\
  \hline
  $H_b$  &    FFT  &    48   &   0.0  &   4.66286 &  periodic \\
  \hline
  $H_z$  &    FFT  &    144    & 0.5   &   12& \\
  \hline
  $q_1$  &    HO   &    35     & 0.0   &  1.0 &  1.0\\
  \hline
  $q_2$  &    HO   &    35     & 0.0   &  1.0 &  1.0\\
  \hline
  $q_3$  &    HO   &    27     & 0.0   &  1.0 &  1.0\\
  \hline
  $q_4$  &    HO   &    27     & 0.0   &  1.0 &  1.0\\
  \hline
  $q_5$  &    HO   &    27     & 0.0   &  1.0 &  1.0\\
  \hline
  $q_6$  &    HO   &    27     & 0.0   &  1.0 &  1.0\\
  \hline
  $q_7$  &    HO   &    27     & 0.0   &  1.0 &  1.0\\
  \hline
  $q_8$  &    HO   &    27     & 0.0   &  1.0 &  1.0\\
  \hline
   {$\!\begin{aligned}
      & q_9 \\
      & \vdots \\
      & q_{69} \\
  \end{aligned}$}
   &   HO   &    25     & 0.0   &  1.0 &  1.0\\
  \hline
    \end{tabularx}
    \label{tab:dvr75D}
\end{table}

Even for the reduced 15D system, the MCTDH calculations are still computationally very expensive. The Multi-Layer (ML)
formulation of MCTDH helps to  decrease the computational costs. Here, instead of directly expanding the SPFs in terms
of the primitive basis, as in  Eq.~(\ref{eq:spf}), several layers of the expansion are introduced: the SPFs that depend on several DoFs are themselves expressed in terms of SPFs of lower dimensionality leading to a very compact expression of the wave functions. For a comprehensive description of ML-MCTDH, we refer the reader to 
Refs. [\!\!\citenum{Wang2003,man08:164116,Vendrell2011}]. 

\begin{figure}
    \centering
    \includegraphics[width=0.9\textwidth]{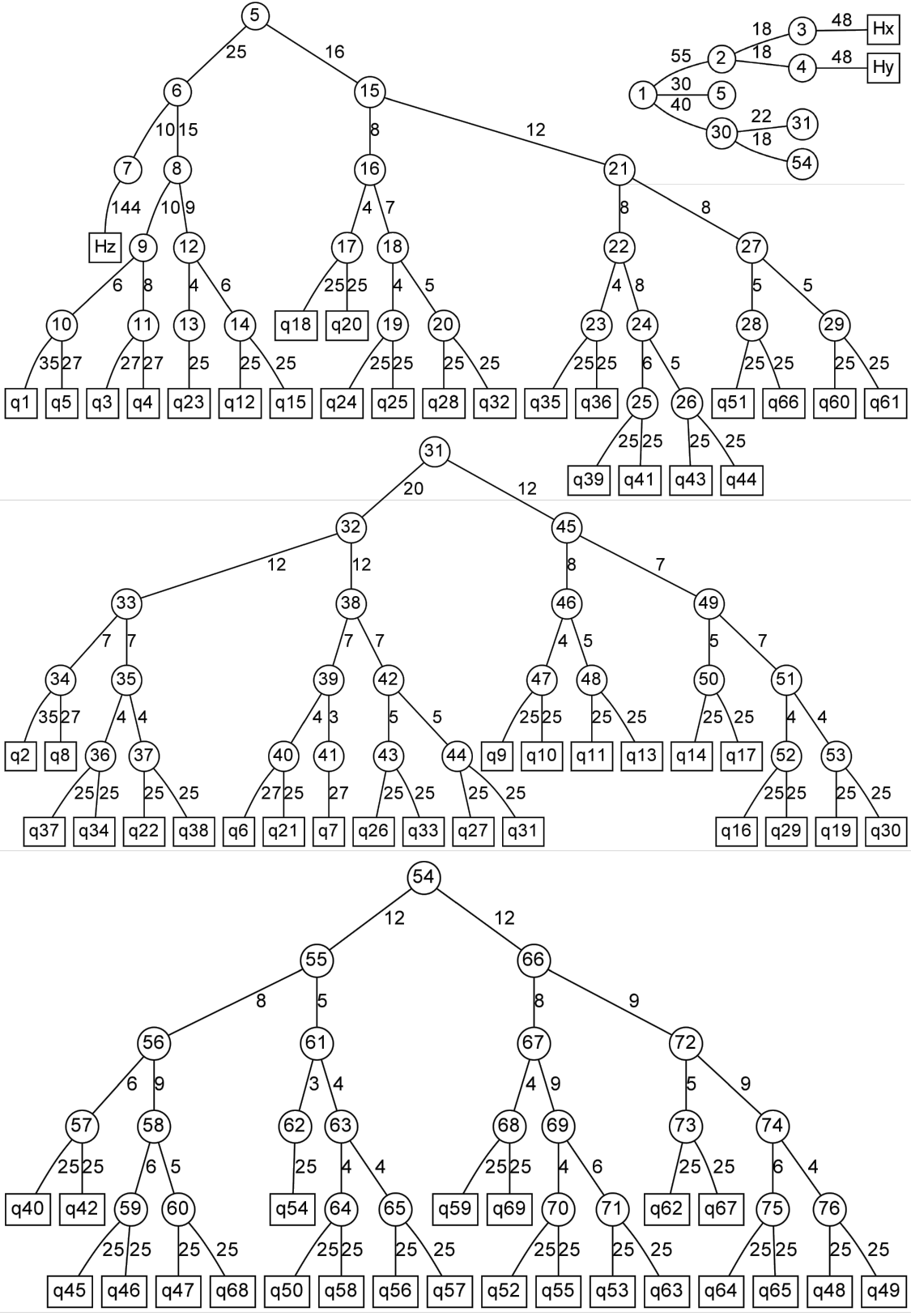}
    \caption{ML-tree for the 75D calculations, the main part is shown on the top left. The construction of node 5, node 31, and node 54 are shown in the figures successively.}
    \label{fig:75MLtree}
\end{figure}

There are many possibilities for the choice of the structure of a
ML-wavefunction. These structures are expressed graphically in a so-called
ML-MCTDH "tree". The choice of a tree can largely influence the speed of
calculation and the complexity of convergence of simulations. Therefore, one should first analyze the correlations between the DOFs in order to use an appropriate ML-tree and accelerate the simulations. We used the DMC samplings when the H atom on the graphene surface to calculate their "correlation matrix" \cite{men23:1144}. The DoFs that are strongly correlated should be combined together. With this approach, the  multi-layer tree is manually constructed and is shown in Fig. \ref{fig:75MLtree}. During the propagations, the population of the lowest occupied time-dependent so-called natural orbitals for each node should be smaller than $10^{-3}$ to achieve good convergence. 

We carefully checked that the lowest occupied time-dependent functions for all nodes in the H atom scattering match the latter condition. We found that the structure of the ML tree displayed in Fig. \ref{fig:75MLtree} is very efficient, so we used it to perform simulations of the H atom scattering from graphene.

\section{\label{sec:results}Results and discussion}

\subsection{Sticking probability}
\begin{table}
\begin{tabular}{c|c|c|c|c|c|c|c|c}
Dimensionality & \multicolumn{4}{c|}{15D} & \multicolumn{4}{c}{75D} \\
\hline
Projectile &  \multicolumn{2}{c|}{H atom} & \multicolumn{2}{c|}{D atom} & \multicolumn{2}{c|}{H atom} & \multicolumn{2}{c}{D atom}\\
\hline
Energy    & 1.96 eV    & 0.96 eV    & 1.96 eV  & 0.96 eV   & 1.96 eV    & 0.96 eV    & 1.96 eV  & 0.96 eV   \\
\hline
cMD  & 0.1\%  & 20.0\%  & 0.9\% & 32.0\% &  1.0\%  & 45.0\%  & 4.3\% & 56.0\% \\
QD  & 0.2\%   & 29.0\%   & 1.6\%  & 39.0\% & 2.0\%   & 56.3\%  & 5.1\% & 67.2\% \\
\end{tabular}
 \caption{Sticking probabilities as functions of the initial conditions.}
     \label{tab:stick}
\end{table}

The sticking probabilities for all cMD and QD simulations with different
collision energies are shown in table \ref{tab:stick}. The sticking probability calculated in the QD simulations and the cMD simulations are close for the initial kinetic energy of 1.96 eV. Compared are the full dimensional system simulation results with our previous 15D simulations\cite{shi23:194102,shi23:059901}, in which the H atom
attacks on the top of one carbon atom. The differences of sticking 
probability for the different collision energies are similar in both 
dimensionalities. A quantum enhancement of sticking probability still
appears and is more evident for low incident energy. However, the
sticking probabilities for same collision energy are larger in the full
dimensional simulations compared with the previous 15D simulation, which
are also shown in the table \ref{tab:stick}. We attribute this larger
sticking to the presence of more surface modes capable of absorbing
more excess energy from the newly formed C-H chemical bond.\par

\subsection{Scattering distributions}
\begin{figure}
    \includegraphics[width=0.8\linewidth]{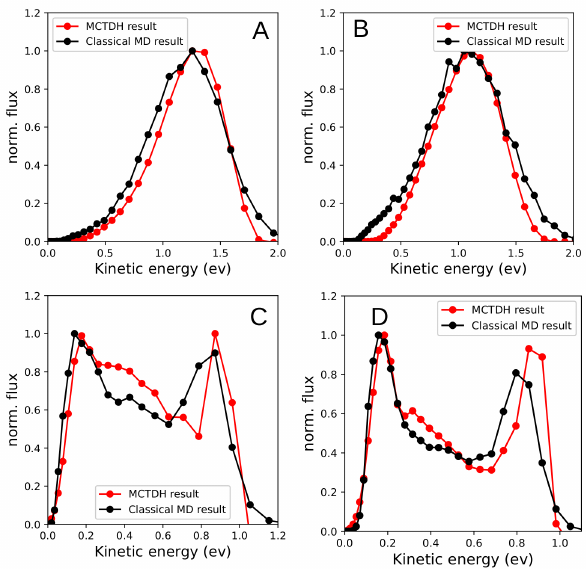}
    \caption{Outgoing kinetic energy distribution for the H atom (A and C) and the D atom (B and D) with the initial kinetic energy of 1.96 $eV$ (A and B) and 0.96 $eV$ (C and D) for cMD simulations (black curves) and for QD simulations (red curves) for 75D simulation with plane wave.}
    \label{fig:eout75D}
\end{figure}

\begin{figure*}[t]
  \includegraphics[width=0.7\linewidth]{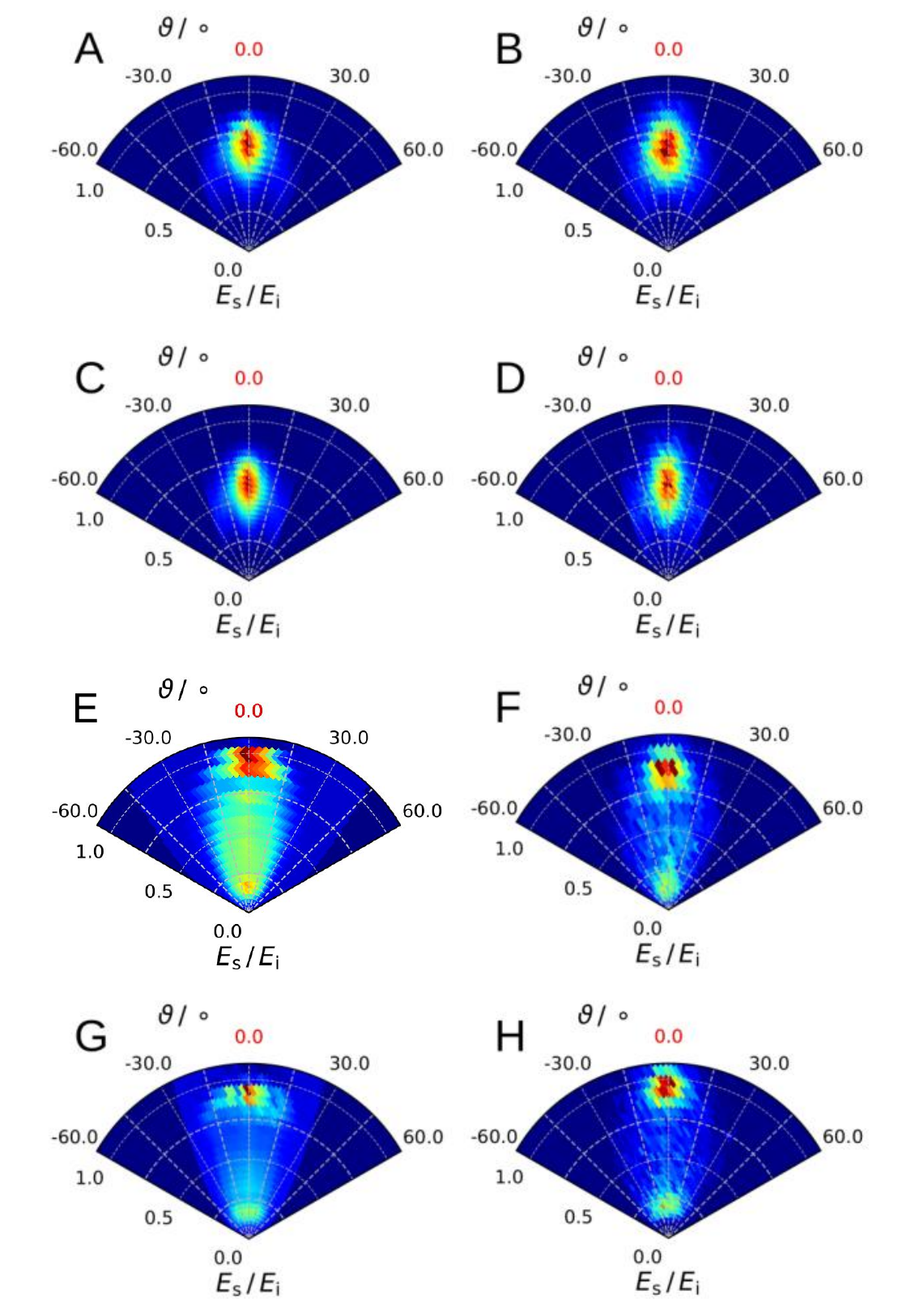}
\caption{2D scattering distribution diagrams for QD simulations (A, C, E, G) and for cMD simulations (B, D, F, H) for the H atom (A, B, E, F) and the D atom (C, D, G, H) with the initial kinetic energy of 1.96 $eV$ (A, B, C, D) and 0.96 $eV$ (E, F, G, H) for 75D simulation with plane wave.}
\label{fig:2dscat}
\end{figure*}

We calculated the H atom outgoing kinetic energy distributions: they
are shown in Figs. \ref{fig:eout75D} and Fig. \ref{fig:2dscat} 
(they look similar to two figures in Ref. \cite{shi23:194102}: this
is a mistake corrected in Ref. \cite{shi23:059901}).  The cMD results
and QD results are different, but the difference is small. Compared
with our 15D  calculations\cite{shi23:194102}, the results of the
initial energy of 1.96 $eV$ are very similar, however, for an incident
energy of 0.96 $eV$, 2 peaks appear in the full dimensional simulations,
one corresponds to an inelastic collision with very low outgoing
kinetic energy. This part is also observed in the 15D simulations.
Another peak corresponds to elastic collisions from a barrier in the 
entrance channel. Because of using a plane wave, the H atom visits
different areas of the surface, and not predominantly the top of one
surface atom.

The scattering distribution of the H atom with the initial conditions of 1.96 $eV$ and 0.96 $eV$ can be compared directly with the experiment results publishe\cite{Exp2022}. For the initial kinetic energy of 1.96 $eV$, both experiments and our simulations lead to a broad outgoing kinetic energy distribution, however the position of the peak in our simulation is slightly larger than that of experiment. This may be explained by a slightly lower chemisorption barrier in the used PES. Potential energy of the used chemisorption barrier is about 0.1 $eV$ smaller than the experimental results. Compared with previous cMD calculations, the outgoing kinetic energy distribution is broader in our new simulations and closer to the experimental results. This is because the zero point energy of the surface was not considered in the previous cMD simulations. Thus, the graphene initial distribution is different. For the initial kinetic energy of 0.96 $eV$, only elastic collision peak is observed in the experimental result and their cMD simulation, this is may due to the different temperature of the graphene surface (remember, we assume a surface temperature of 0 K). 

\subsection{Collective modes analysis}

As we worked with a rather large and periodic graphene cell, the calculated normal modes, which are collective modes, resemble phonons of a graphene surface. We followed how the collective modes react during the collision and how the energy is redistributed during the collision, {in order to gain insight into the role of phonons of a graphene surface}. The results are given for an H atom with the initial kinetic energies of 0.96 $eV$ and 1.96 $eV$.  

\begin{figure*}
    \includegraphics[width=0.7\linewidth]{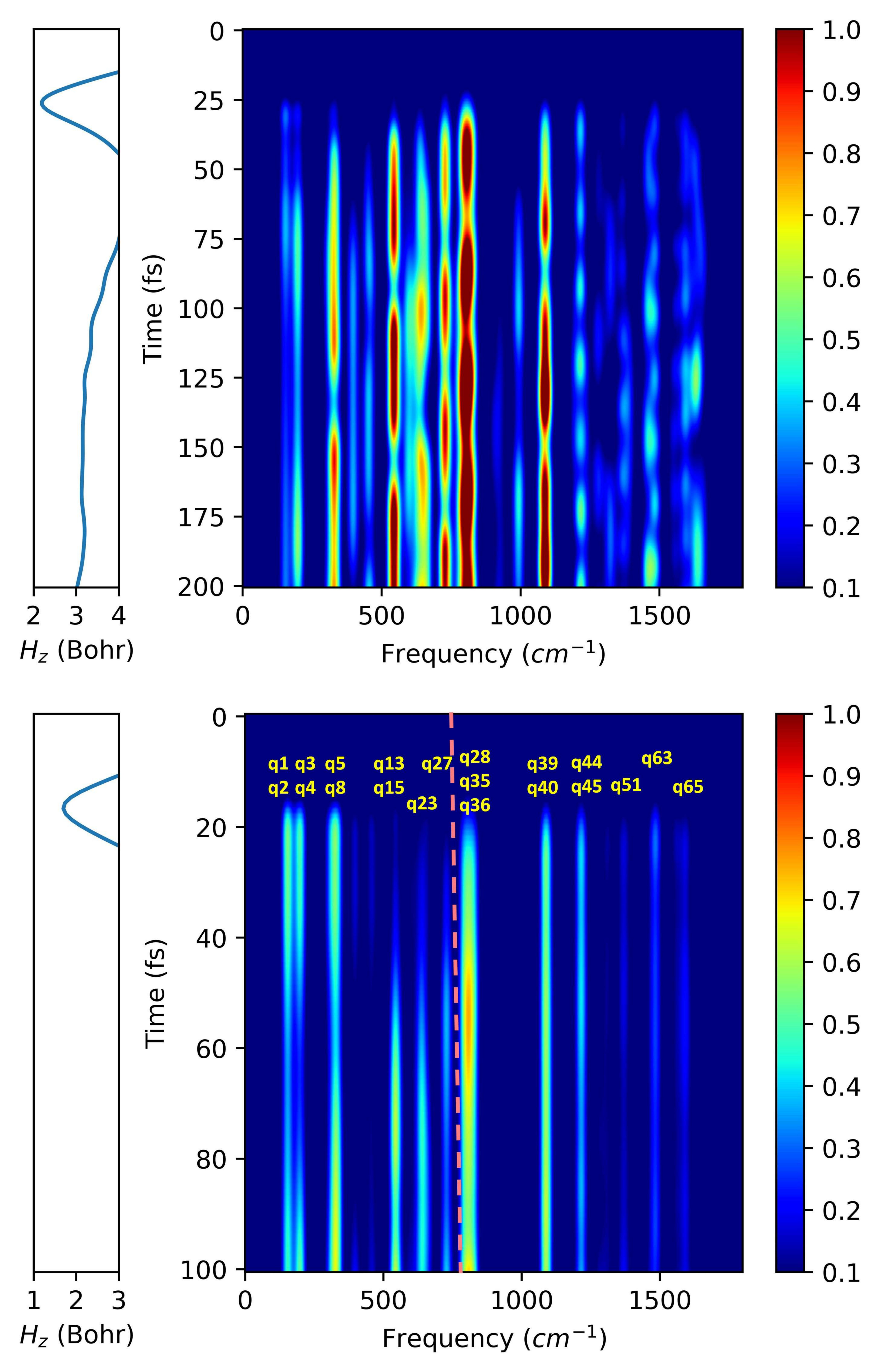}
    \caption{Populations of graphene collective modes as a function of time for an initial kinetic energy of 0.96 $eV$ (top) and 1.96 $eV$ (bottom). The X-axis is the frequency of the collective modes, the Y-axis is the time of simulation. The position of the H atom during the simulation is represented by the blue curve on the left, with its X-axis representing the distance between the H atom and the graphene surface. The colormap represents the populations of the excited collective modes.}
    \label{fig:2dphonon}
\end{figure*}

The results are displayed in Fig. \ref{fig:2dphonon}. The color map shows the population of the excited collective modes. We also gave a label to the normal modes that is given in the figure corresponding to the dynamics with an incident energy of 1.96 $eV$.
\begin{figure*}
    \includegraphics[width=0.45\linewidth]{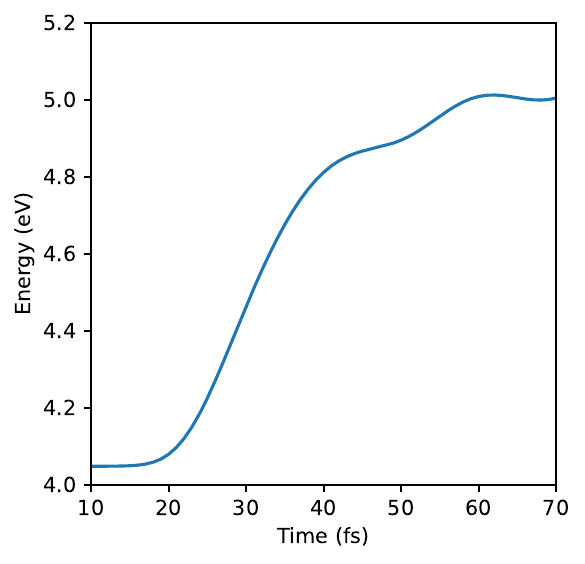}
    \includegraphics[width=0.45\linewidth]{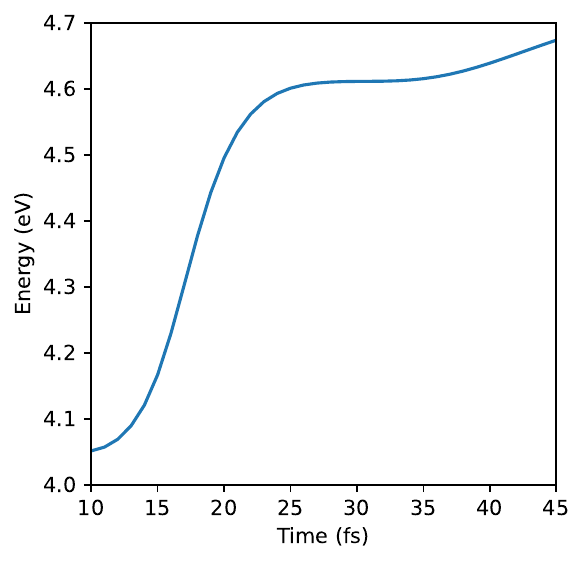}
    \caption{Graphene surface total energy during the collision for the 75D  simulations with an incident energy of 0.96 $eV$ (left) and with an incident energy of 1.96 $eV$ (right).}
    \label{fig:slabene}
\end{figure*}

Several key observations can be made. Comparing the results of the incident energy of 0.96 $eV$ ans 1.96 $eV$, the excited modes are almost the same, there are some other modes different than the modes indicated in Fig. \ref{fig:2dphonon} that are slightly excited in the case of an initial kinetic energy of 0.96 $eV$. Besides, there is more excitation for the incident energy of 0.96 $eV$. The differences come
from the fact that  the graphene surface absorbs more energy in the case of 0.96 $eV$, in agreement with Fig. \ref{fig:slabene}. 

Collective modes begin to be excited when the H atom start to reach the
surface, with some collective modes excited earlier than others. Many
collective modes are involved in the collision at the onset of
interaction between the H atom and the C atom.
The excited modes can be classified by separating them with the pink 
dashed line in the figure corresponding to a given frequency. All
excited modes with lower frequencies than the one associated to this
pink line correspond to out-of-plane modes, the modes with higher
frequencies are in-plane modes. The populations of the excited collective modes undergo dynamic changes during the simulation. For example, for the incident kinetic energy of 1.96 $eV$, the population of the collective mode $q_5$ decreases at approximately 50 fs, while the population of the collective mode $q_{13}$ increases. This phenomenon is much more evident for the incident kinetic energy of 0.96 $eV$. This might suggest that these mode are very correlated due to Fermi-like resonances.

The collision between the H atom and the graphene surface has a specific behavior. 
The impacted atom does not recoil towards the projectile. One does not observe an oscillation of energy for the attacked object: see Figs. \ref{fig:slabene}. {This figure depicts how the total energy of the graphene surface changes during the collision. The surface energy increases quickly during the collision, and then remains almost constant.} The surface energy doesn't decrease during the collision. 

Furthermore, it is worth noting that like in numerous prior studies 
\cite{Bonfanti2015,Bonfanti_2018}, our work has has highlighted the 
major role of the out-of-plane collective modes with frequencies 
below 800 $cm^{-1}$ that allow the creation of a CH bond. They 
play a role similar to the normal mode $q_1$ and $q_4$ in the 15D
system study in our previous study. The in-plane collective modes
also play a very important role since the absorb the excess of energy
of the H atom. They play a role similar to the normal mode $q_{10}$,
the breathing movement of the first shell, in the 15D system study.
The definition of the normal modes with a population larger than 0.1
are given in the appendix. There are 20 normal modes involved in
collision, 10 out-of-plane modes, and 10 in-plane modes.

\section{Conclusion}
After thoroughly analyzing the results of our comprehensive cMD and QD
simulations under strictly comparable initial conditions and two
collision energies, we have drawn several significant conclusions.
First, the coherence between the cMD and QD simulations, shown both
in the present work and in our prior 15D
simulations\cite{shi23:194102,shi23:059901}, validates the robustness
and accuracy of our simulation methodologies. This is very important
in view of the very high complexity of doing 75D quantum simulations
starting form an {\it ab initio} potential energy surface. 

Second, by directly comparing our simulation results with experimental
data, we have identified key factors contributing to discrepancies
observed between previous\cite{shi23:194102} cMD simulations and
experimental observations. Specifically, we found that inaccuracies
in the representation of the chemisorption barrier and vibrational
ground state distribution are likely to be the main source of these
discrepancies. 

Furthermore, our in-depth analysis of the collective modes present on the graphene surface during the collision process provides valuable insights into the underlying mechanisms governing energy dispersion. Specifically, we observed that in-plane collective modes play a crucial role in mediating energy dissipation, which has important implications for understanding and controlling energy transfer processes.

In summary, our work not only demonstrates the coherence and reliability of our simulation techniques but also identifies key factors that can be targeted for improvement in future studies. The quantum effects are still not very strong in those simulations.
Working with different initial conditions, low incident energy and non-perpendicular incident angles will very likely lead to situations where the quantum effects are stronger, which could explain some experimental findings.  Work in this direction is in progress.

\section{Acknowledgement}
The authors thank the computational platforms IDRIS HPE Jean Zay (project A0110812942) and Ruche Mesocentre and the CNRS International Research Network (IRN) “MCTDH” for financial support.  A.K.\  acknowledges support from the European Research Council (ERC) under the European Union’s Horizon 2020 research and innovation programme (grant agreement no. 833404). 
\newpage
\appendix
\section{Normal modes}
\label{apx:a}
In this appendix, we give the definition of the normal modes. 

The molecular Hamiltonian in Cartesian coordinates around the equilibrium position $\mathbf{x_0}$ can be written by Taylor formula as:
\begin{equation}
    \hat{H}(\mathbf{x})=\sum_i - \frac{\hbar^2}{2m_i^2}\frac{\partial ^ 2}{\partial x_i^2}+V(\mathbf{x_0})+\sum_{ij}\frac{1}{2}\frac{\partial^2 V}{\partial x_{0i}\partial x_{0j}}(x_i-x_{0i})(x_j-x_{0j})+\dots \nonumber
\end{equation}
because the term $\sum_i\frac{\partial V}{\partial x_{0i}}(x_i-x_{0i})$
vanishes at equilibrium position. By neglecting higher order of the
Taylor series and using mass weighted coordinates, the Hamiltonian can be reformulated as :
\begin{equation}
    \hat{H}(\mathbf{\Tilde{x}})=\sum_i-\frac{\hbar^2}{2}\frac{\partial ^ 2}{\partial \Tilde{x}_i^2}+V(\mathbf{x_0})+\sum_{ij}\frac{1}{2}\frac{1}{\sqrt{m_im_j}}\frac{\partial^2 V}{\partial x_{0i}\partial x_{0j}}\Tilde{x}_i\Tilde{x}_j, \nonumber
\end{equation}
where $\Tilde{x}_i= \sqrt{m_i}(x_i-x_{0i})$, and the term $\frac{1}{\sqrt{m_im_j}}\frac{\partial^2 V}{\partial x_{0i}\partial x_{0j}}$ 
is the mass weighted Hessian matrix for the potential energy around
the equilibrium position. By diagonalization of this matrix, one 
obtains a transformation matrix $\mathbf{D}$ leading to the the normal
modes defined as linear combinations of the Cartesian coordinates :
\begin{equation}
    x_i=\sum_jD_{ij} q_j, \nonumber
\end{equation}
where the $q_j$ is the normal mode. And a set of eigenvalues
$\mathbf{\omega}$. The Hamiltonian in normal coordinates  reads
(we neglected the so-called $\pi \pi$ term):
\begin{equation}
    \label{eq:A1}
    \hat{H}(\mathbf{q})=\sum_i-\frac{\hbar^2}{2}\frac{\partial^2}{\partial q_i^2}+V(\mathbf{x_0})+\sum_i\frac{1}{2}\omega_i^2q_i^2.
\end{equation}
The most important normal coordinates are shown in Fig. \ref{fig:2dphonon}.

There are three vanishing normal mode frequencies, we call them 
$\omega_{70}$, $\omega_{71}$, and $\omega_{72}$. The corresponding
motions are translations of center of mass of the whole surface.
As the center of mass is kept fixed, the corresponding coordinates,
$q_{70}$, $q_{72}$, and $q_{72}$, are set to zero. This can be
accomplished by letting the index $i$ in Eq. \ref{eq:A1} run
form 1 to 69 only.

\begin{figure}
    \centering
    \includegraphics[width=0.4\linewidth]{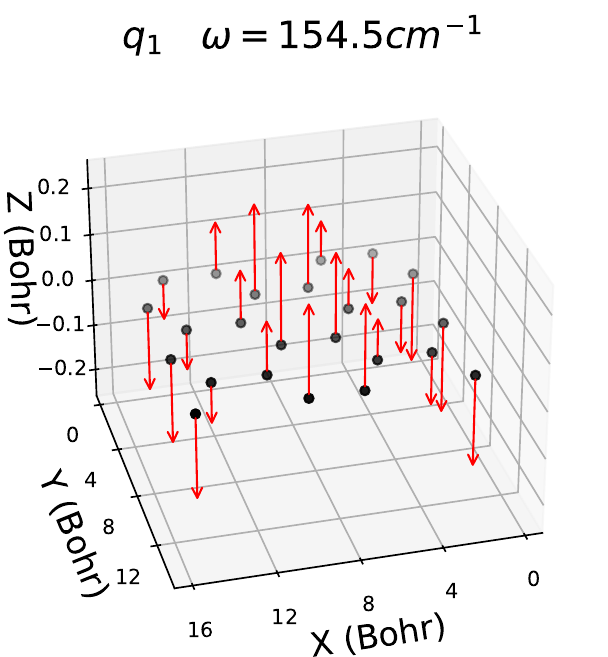}
    \includegraphics[width=0.4\linewidth]{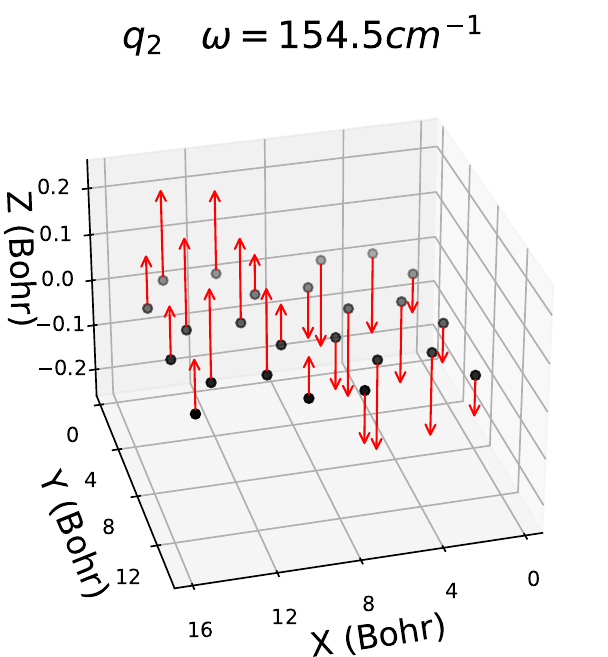}
    \includegraphics[width=0.4\linewidth]{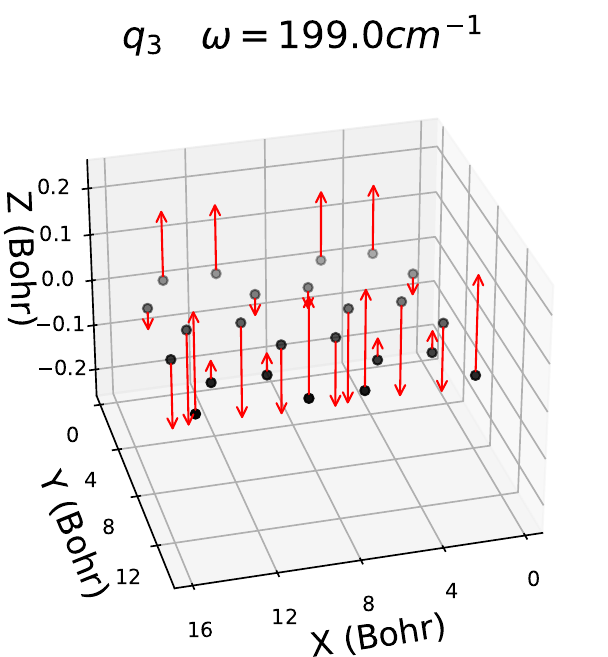}
    \includegraphics[width=0.4\linewidth]{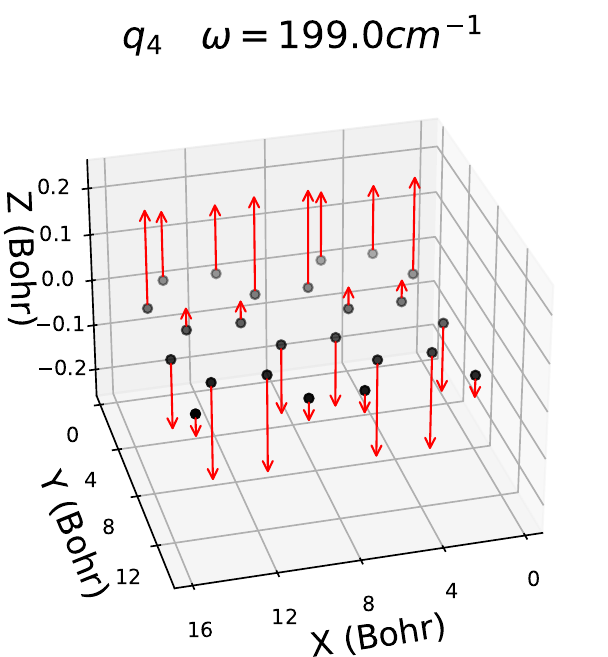}
\end{figure}
\begin{figure}
    \centering
    \includegraphics[width=0.4\linewidth]{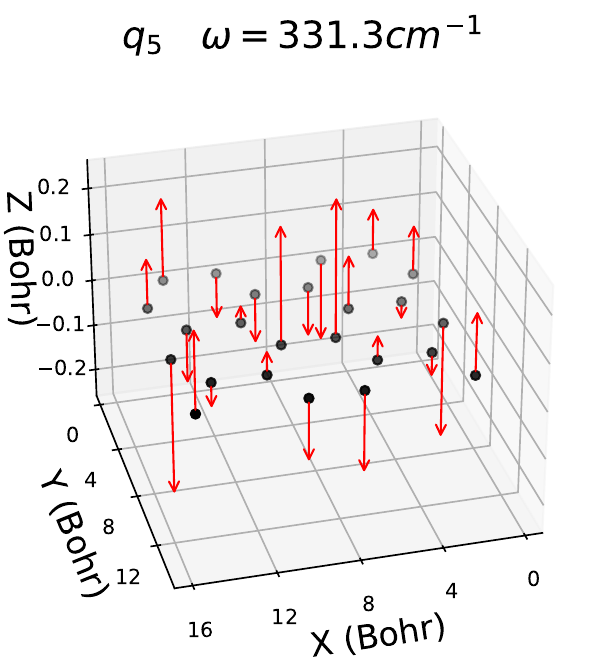}
    \includegraphics[width=0.4\linewidth]{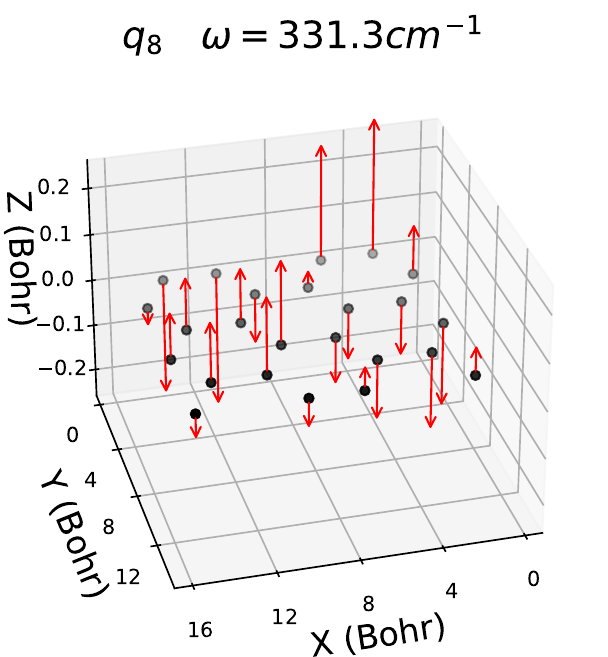}
    \includegraphics[width=0.4\linewidth]{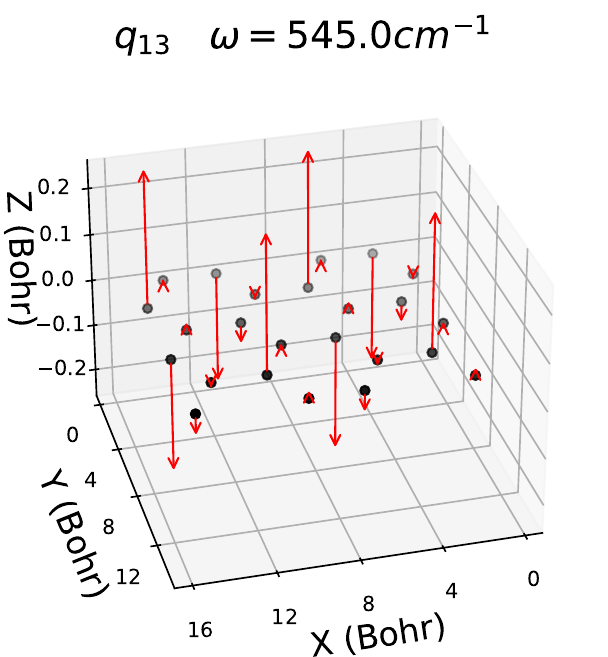}
    \includegraphics[width=0.4\linewidth]{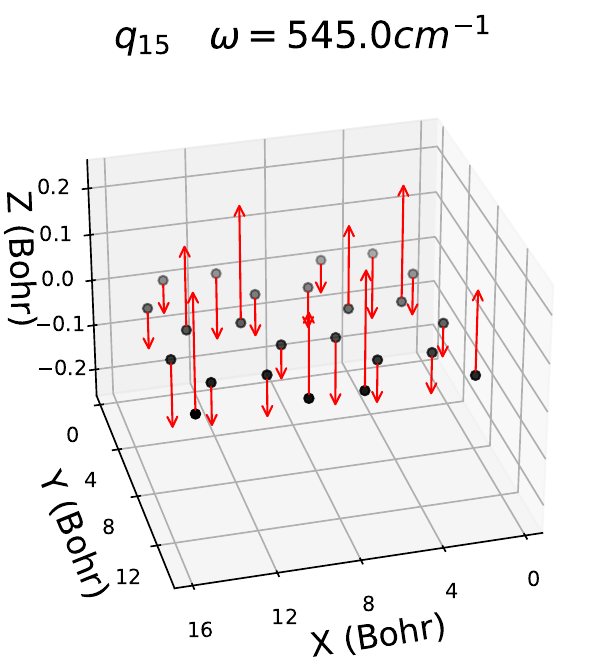}
    \includegraphics[width=0.4\linewidth]{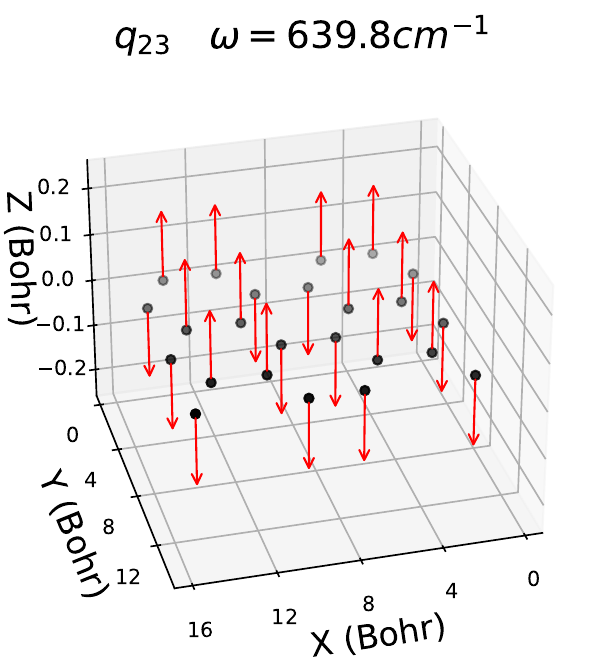}
    \includegraphics[width=0.4\linewidth]{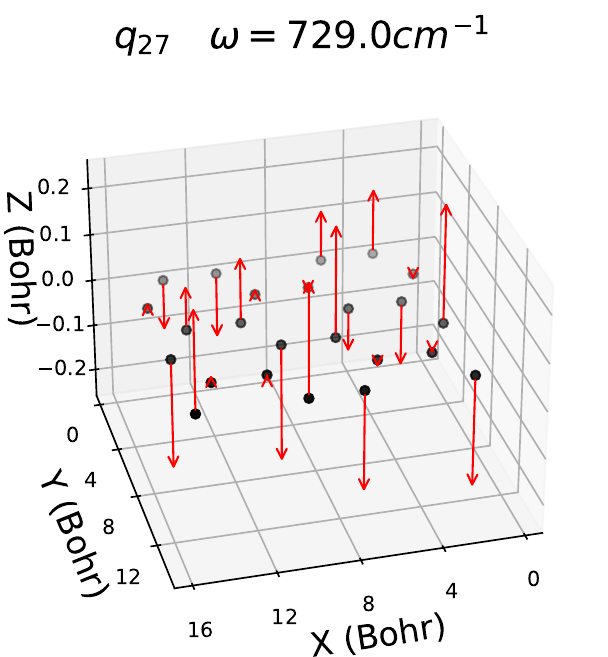}
    \caption{The 10 out of plane normal modes excited during the collision, with the label and frequency indicated on the top of each mode. The position of the C atoms are indicated by the black dots. The gray scale of dots is the Y value of C atom position. The movement are shown by the red arrows. }
\end{figure}
\begin{figure}
    \centering
    \includegraphics[width=0.4\linewidth]{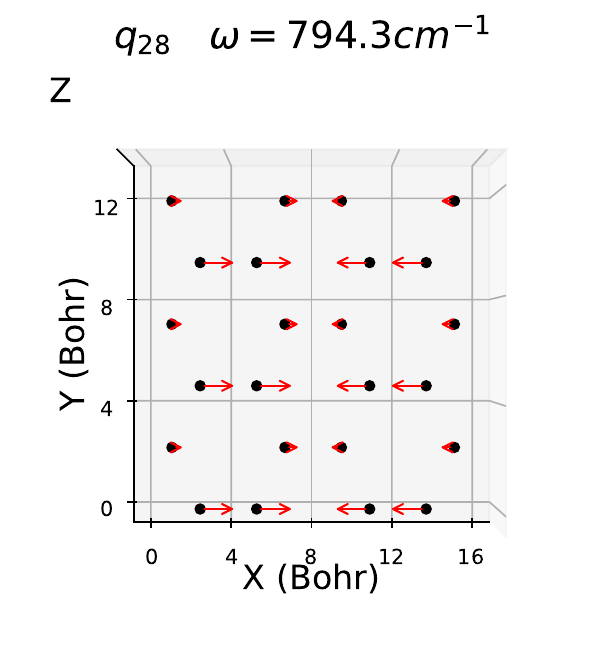}
    \includegraphics[width=0.4\linewidth]{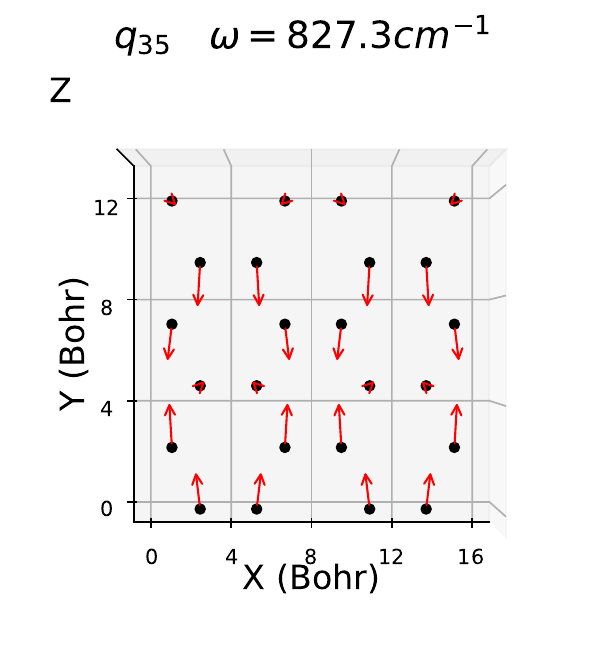}
    \includegraphics[width=0.4\linewidth]{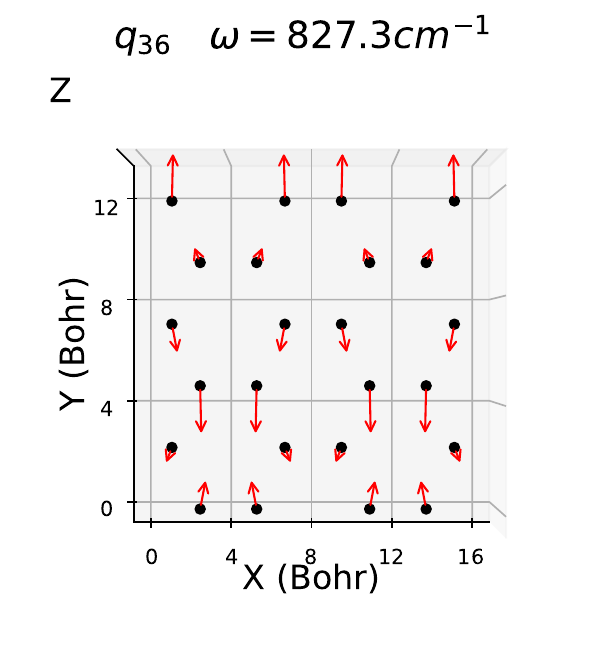}
    \includegraphics[width=0.4\linewidth]{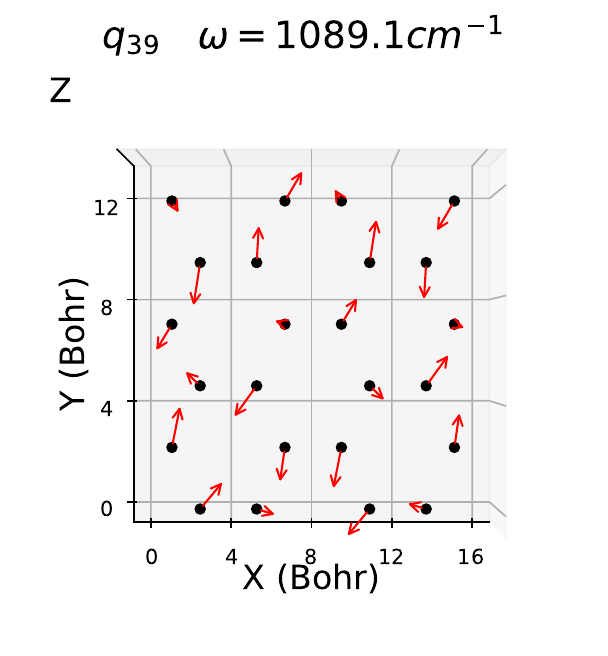}
    \includegraphics[width=0.4\linewidth]{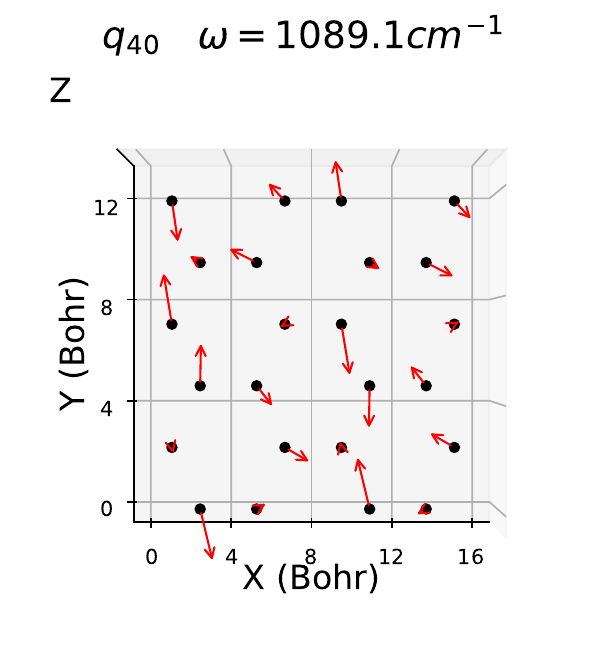}
    \includegraphics[width=0.4\linewidth]{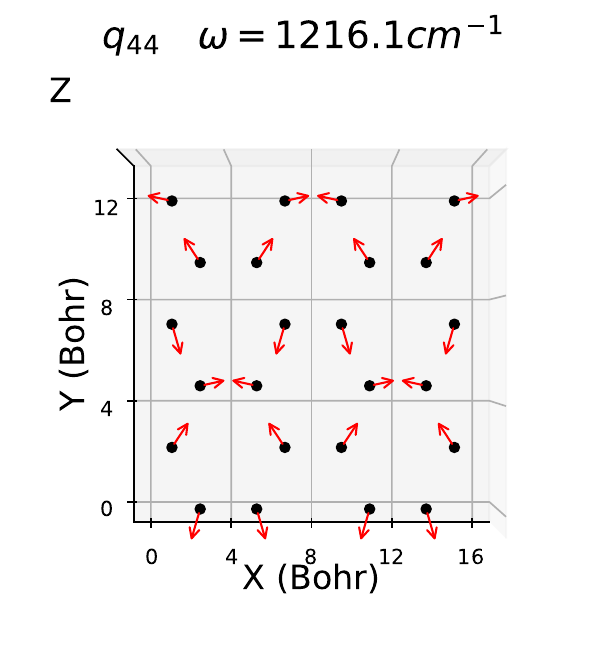}
\end{figure}
\begin{figure}
    \centering
    \includegraphics[width=0.4\linewidth]{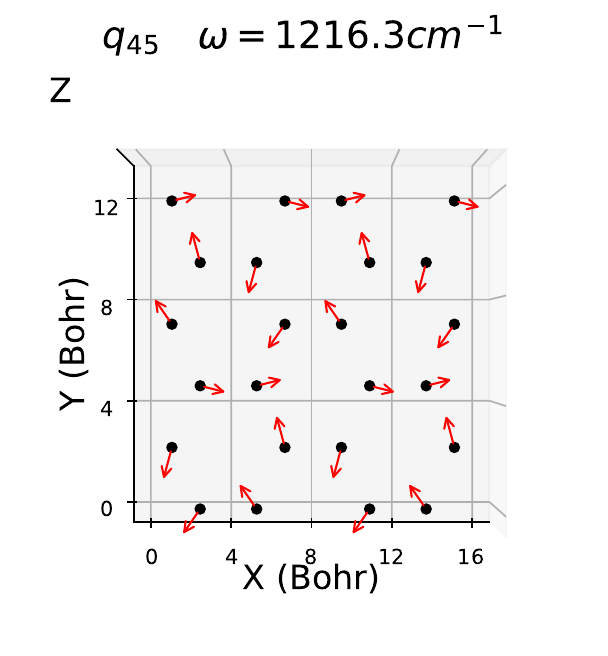}
    \includegraphics[width=0.4\linewidth]{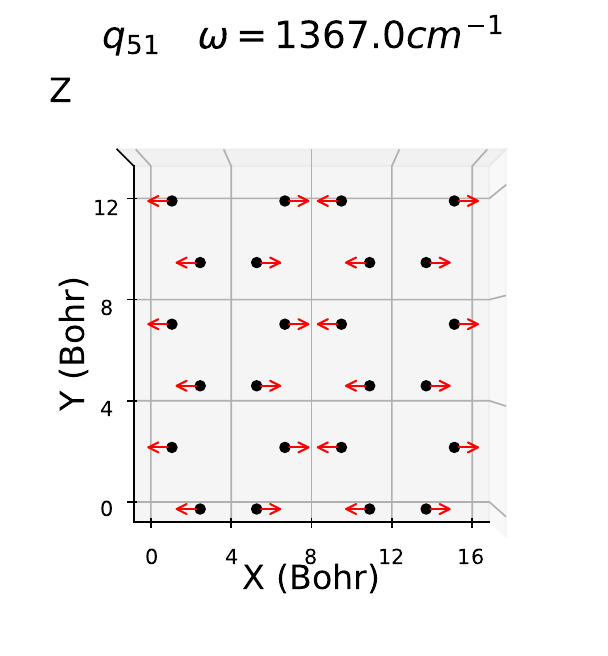}
    \includegraphics[width=0.4\linewidth]{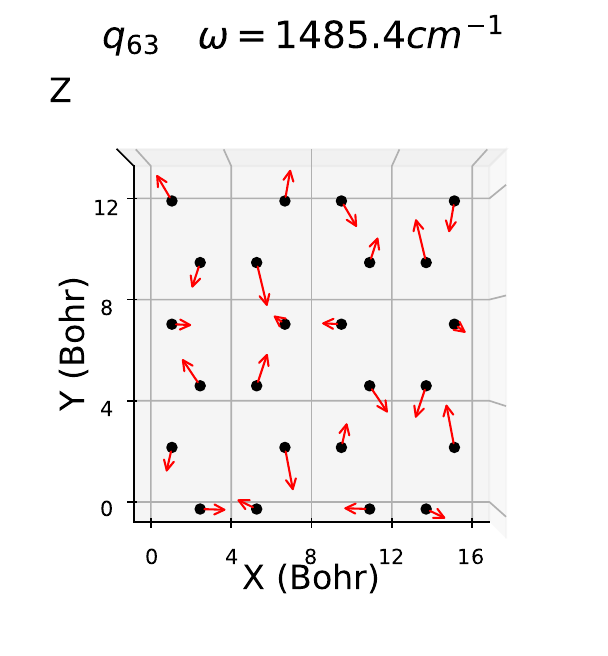}
    \includegraphics[width=0.4\linewidth]{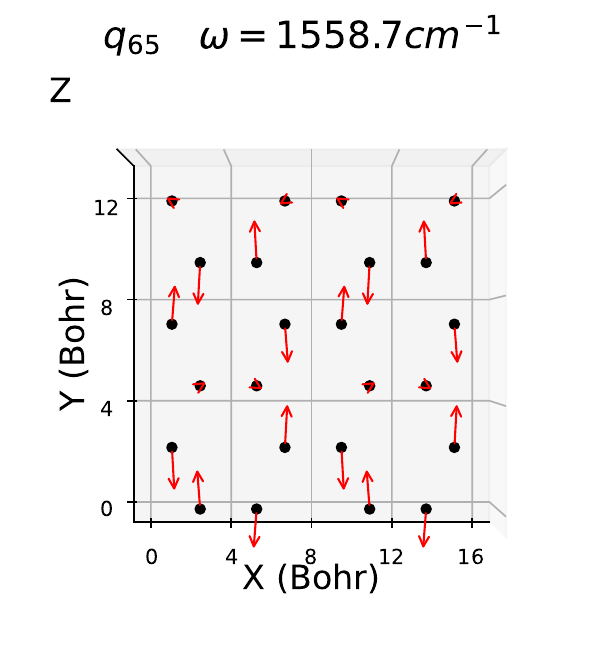}
    \caption{The 10 in plane normal modes excited during the collision, with the label and frequency indicated on the top of each mode. The position of the C atoms are indicated by the black dots. The movement are shown by the red arrows.}
\end{figure}

\clearpage

\nocite{*}

\begin{thebibliography}{33}%
\makeatletter
\providecommand \@ifxundefined [1]{%
 \@ifx{#1\undefined}
}%
\providecommand \@ifnum [1]{%
 \ifnum #1\expandafter \@firstoftwo
 \else \expandafter \@secondoftwo
 \fi
}%
\providecommand \@ifx [1]{%
 \ifx #1\expandafter \@firstoftwo
 \else \expandafter \@secondoftwo
 \fi
}%
\providecommand \natexlab [1]{#1}%
\providecommand \enquote  [1]{``#1''}%
\providecommand \bibnamefont  [1]{#1}%
\providecommand \bibfnamefont [1]{#1}%
\providecommand \citenamefont [1]{#1}%
\providecommand \href@noop [0]{\@secondoftwo}%
\providecommand \href [0]{\begingroup \@sanitize@url \@href}%
\providecommand \@href[1]{\@@startlink{#1}\@@href}%
\providecommand \@@href[1]{\endgroup#1\@@endlink}%
\providecommand \@sanitize@url [0]{\catcode `\\12\catcode `\$12\catcode
  `\&12\catcode `\#12\catcode `\^12\catcode `\_12\catcode `\%12\relax}%
\providecommand \@@startlink[1]{}%
\providecommand \@@endlink[0]{}%
\providecommand \url  [0]{\begingroup\@sanitize@url \@url }%
\providecommand \@url [1]{\endgroup\@href {#1}{\urlprefix }}%
\providecommand \urlprefix  [0]{URL }%
\providecommand \Eprint [0]{\href }%
\providecommand \doibase [0]{https://doi.org/}%
\providecommand \selectlanguage [0]{\@gobble}%
\providecommand \bibinfo  [0]{\@secondoftwo}%
\providecommand \bibfield  [0]{\@secondoftwo}%
\providecommand \translation [1]{[#1]}%
\providecommand \BibitemOpen [0]{}%
\providecommand \bibitemStop [0]{}%
\providecommand \bibitemNoStop [0]{.\EOS\space}%
\providecommand \EOS [0]{\spacefactor3000\relax}%
\providecommand \BibitemShut  [1]{\csname bibitem#1\endcsname}%
\let\auto@bib@innerbib\@empty
\bibitem [{\citenamefont {Fang}\ \emph {et~al.}(2020)\citenamefont {Fang},
  \citenamefont {Zhang}, \citenamefont {Chen},\ and\ \citenamefont
  {Weng}}]{Fang2020}%
  \BibitemOpen
  \bibfield  {author} {\bibinfo {author} {\bibfnamefont {C.}~\bibnamefont
  {Fang}}, \bibinfo {author} {\bibfnamefont {J.}~\bibnamefont {Zhang}},
  \bibinfo {author} {\bibfnamefont {X.}~\bibnamefont {Chen}},\ and\ \bibinfo
  {author} {\bibfnamefont {G.~J.}\ \bibnamefont {Weng}},\ }\bibfield  {title}
  {\enquote {\bibinfo {title} {Calculating the electrical conductivity of
  graphene nanoplatelet polymer composites by a monte carlo method},}\ }\href
  {https://doi.org/10.3390/nano10061129} {\bibfield  {journal} {\bibinfo
  {journal} {Nanomaterials}\ }\textbf {\bibinfo {volume} {10}},\ \bibinfo
  {pages} {1--15} (\bibinfo {year} {2020})}\BibitemShut {NoStop}%
\bibitem [{\citenamefont {Burnett}\ \emph {et~al.}(1999)\citenamefont
  {Burnett}, \citenamefont {Julienne}, \citenamefont {Lett}, \citenamefont
  {Tiesinga}, \citenamefont {Williams}, \citenamefont {Bethlem}, \citenamefont
  {Berden}, \citenamefont {van Roij}, \citenamefont {Hornekaer}, \citenamefont
  {Baurichter}, \citenamefont {Petrunin}, \citenamefont {Field},\ and\
  \citenamefont {Luntz}}]{Burnett1999}%
  \BibitemOpen
  \bibfield  {author} {\bibinfo {author} {\bibfnamefont {K.}~\bibnamefont
  {Burnett}}, \bibinfo {author} {\bibfnamefont {P.~S.}\ \bibnamefont
  {Julienne}}, \bibinfo {author} {\bibfnamefont {P.~D.}\ \bibnamefont {Lett}},
  \bibinfo {author} {\bibfnamefont {E.}~\bibnamefont {Tiesinga}}, \bibinfo
  {author} {\bibfnamefont {C.~J.}\ \bibnamefont {Williams}}, \bibinfo {author}
  {\bibfnamefont {H.~L.}\ \bibnamefont {Bethlem}}, \bibinfo {author}
  {\bibfnamefont {G.}~\bibnamefont {Berden}}, \bibinfo {author} {\bibfnamefont
  {A.~J.~A.}\ \bibnamefont {van Roij}}, \bibinfo {author} {\bibfnamefont
  {L.}~\bibnamefont {Hornekaer}}, \bibinfo {author} {\bibfnamefont
  {A.}~\bibnamefont {Baurichter}}, \bibinfo {author} {\bibfnamefont {V.~V.}\
  \bibnamefont {Petrunin}}, \bibinfo {author} {\bibfnamefont {D.}~\bibnamefont
  {Field}},\ and\ \bibinfo {author} {\bibfnamefont {A.~C.}\ \bibnamefont
  {Luntz}},\ }\bibfield  {title} {\enquote {\bibinfo {title} {10. w.
  ketterle},}\ }\href {www.sciencemag.org} {\bibfield  {journal} {\bibinfo
  {journal} {Adv. Atom. Mol. Opt. Phys}\ }\textbf {\bibinfo {volume} {71}},\
  \bibinfo {pages} {6455} (\bibinfo {year} {1999})}\BibitemShut {NoStop}%
\bibitem [{\citenamefont {Novoselov}\ \emph {et~al.}(2012)\citenamefont
  {Novoselov}, \citenamefont {Fal'Ko}, \citenamefont {Colombo}, \citenamefont
  {Gellert}, \citenamefont {Schwab},\ and\ \citenamefont
  {Kim}}]{Novoselov2012}%
  \BibitemOpen
  \bibfield  {author} {\bibinfo {author} {\bibfnamefont {K.~S.}\ \bibnamefont
  {Novoselov}}, \bibinfo {author} {\bibfnamefont {V.~I.}\ \bibnamefont
  {Fal'Ko}}, \bibinfo {author} {\bibfnamefont {L.}~\bibnamefont {Colombo}},
  \bibinfo {author} {\bibfnamefont {P.~R.}\ \bibnamefont {Gellert}}, \bibinfo
  {author} {\bibfnamefont {M.~G.}\ \bibnamefont {Schwab}},\ and\ \bibinfo
  {author} {\bibfnamefont {K.}~\bibnamefont {Kim}},\ }\bibfield  {title}
  {\enquote {\bibinfo {title} {A roadmap for graphene},}\ }\href
  {https://doi.org/10.1038/nature11458} {\bibfield  {journal} {\bibinfo
  {journal} {Nature}\ }\textbf {\bibinfo {volume} {490}},\ \bibinfo {pages}
  {192--200} (\bibinfo {year} {2012})}\BibitemShut {NoStop}%
\bibitem [{\citenamefont {Schlapbach}\ and\ \citenamefont
  {Züttel}(2001)}]{Schlapbach2001}%
  \BibitemOpen
  \bibfield  {author} {\bibinfo {author} {\bibfnamefont {L.}~\bibnamefont
  {Schlapbach}}\ and\ \bibinfo {author} {\bibfnamefont {A.}~\bibnamefont
  {Züttel}},\ }\enquote {\bibinfo {title} {Hydrogen-storage materials for
  mobile applications},}\ in\ \href
  {https://doi.org/10.1142/9789814317665_0038} {\emph {\bibinfo {booktitle}
  {Materials for Sustainable Energy}}}\ (\bibinfo  {publisher} {World
  Scientific},\ \bibinfo {year} {2001})\ pp.\ \bibinfo {pages}
  {265--270}\BibitemShut {NoStop}%
\bibitem [{\citenamefont {Lonkar}, \citenamefont {Deshmukh},\ and\
  \citenamefont {Abdala}(2015)}]{Lonkar2015}%
  \BibitemOpen
  \bibfield  {author} {\bibinfo {author} {\bibfnamefont {S.~P.}\ \bibnamefont
  {Lonkar}}, \bibinfo {author} {\bibfnamefont {Y.~S.}\ \bibnamefont
  {Deshmukh}},\ and\ \bibinfo {author} {\bibfnamefont {A.~A.}\ \bibnamefont
  {Abdala}},\ }\bibfield  {title} {\enquote {\bibinfo {title} {Recent advances
  in chemical modifications of graphene},}\ }\href
  {https://doi.org/10.1007/s12274-014-0622-9} {\bibfield  {journal} {\bibinfo
  {journal} {Nano Research}\ }\textbf {\bibinfo {volume} {8}},\ \bibinfo
  {pages} {1039--1074} (\bibinfo {year} {2015})}\BibitemShut {NoStop}%
\bibitem [{\citenamefont {Balog}\ \emph {et~al.}(2010)\citenamefont {Balog},
  \citenamefont {Jørgensen}, \citenamefont {Nilsson}, \citenamefont
  {Andersen}, \citenamefont {Rienks}, \citenamefont {Bianchi}, \citenamefont
  {Fanetti}, \citenamefont {Lægsgaard}, \citenamefont {Baraldi}, \citenamefont
  {Lizzit}, \citenamefont {Sljivancanin}, \citenamefont {Besenbacher},
  \citenamefont {Hammer}, \citenamefont {Pedersen}, \citenamefont {Hofmann},\
  and\ \citenamefont {Hornekær}}]{Balog2010}%
  \BibitemOpen
  \bibfield  {author} {\bibinfo {author} {\bibfnamefont {R.}~\bibnamefont
  {Balog}}, \bibinfo {author} {\bibfnamefont {B.}~\bibnamefont {Jørgensen}},
  \bibinfo {author} {\bibfnamefont {L.}~\bibnamefont {Nilsson}}, \bibinfo
  {author} {\bibfnamefont {M.}~\bibnamefont {Andersen}}, \bibinfo {author}
  {\bibfnamefont {E.}~\bibnamefont {Rienks}}, \bibinfo {author} {\bibfnamefont
  {M.}~\bibnamefont {Bianchi}}, \bibinfo {author} {\bibfnamefont
  {M.}~\bibnamefont {Fanetti}}, \bibinfo {author} {\bibfnamefont
  {E.}~\bibnamefont {Lægsgaard}}, \bibinfo {author} {\bibfnamefont
  {A.}~\bibnamefont {Baraldi}}, \bibinfo {author} {\bibfnamefont
  {S.}~\bibnamefont {Lizzit}}, \bibinfo {author} {\bibfnamefont
  {Z.}~\bibnamefont {Sljivancanin}}, \bibinfo {author} {\bibfnamefont
  {F.}~\bibnamefont {Besenbacher}}, \bibinfo {author} {\bibfnamefont
  {B.}~\bibnamefont {Hammer}}, \bibinfo {author} {\bibfnamefont {T.~G.}\
  \bibnamefont {Pedersen}}, \bibinfo {author} {\bibfnamefont {P.}~\bibnamefont
  {Hofmann}},\ and\ \bibinfo {author} {\bibfnamefont {L.}~\bibnamefont
  {Hornekær}},\ }\bibfield  {title} {\enquote {\bibinfo {title} {Bandgap
  opening in graphene induced by patterned hydrogen adsorption},}\ }\href
  {https://doi.org/10.1038/nmat2710} {\bibfield  {journal} {\bibinfo  {journal}
  {Nature Materials}\ }\textbf {\bibinfo {volume} {9}},\ \bibinfo {pages}
  {315--319} (\bibinfo {year} {2010})}\BibitemShut {NoStop}%
\bibitem [{\citenamefont {Jiang}\ \emph {et~al.}(2019)\citenamefont {Jiang},
  \citenamefont {Kammler}, \citenamefont {Ding}, \citenamefont {Dorenkamp},
  \citenamefont {Manby}, \citenamefont {Wodtke}, \citenamefont {Miller},
  \citenamefont {Kandratsenka},\ and\ \citenamefont {Bünermann}}]{Jiang2019}%
  \BibitemOpen
  \bibfield  {author} {\bibinfo {author} {\bibfnamefont {H.}~\bibnamefont
  {Jiang}}, \bibinfo {author} {\bibfnamefont {M.}~\bibnamefont {Kammler}},
  \bibinfo {author} {\bibfnamefont {F.}~\bibnamefont {Ding}}, \bibinfo {author}
  {\bibfnamefont {Y.}~\bibnamefont {Dorenkamp}}, \bibinfo {author}
  {\bibfnamefont {F.~R.}\ \bibnamefont {Manby}}, \bibinfo {author}
  {\bibfnamefont {A.~M.}\ \bibnamefont {Wodtke}}, \bibinfo {author}
  {\bibfnamefont {T.~F.}\ \bibnamefont {Miller}}, \bibinfo {author}
  {\bibfnamefont {A.}~\bibnamefont {Kandratsenka}},\ and\ \bibinfo {author}
  {\bibfnamefont {O.}~\bibnamefont {Bünermann}},\ }\bibfield  {title}
  {\enquote {\bibinfo {title} {Imaging covalent bond formation by h atom
  scattering from graphene},}\ }\href {https://doi.org/10.1126/science.aaw6378}
  {\bibfield  {journal} {\bibinfo  {journal} {Science}\ }\textbf {\bibinfo
  {volume} {364}},\ \bibinfo {pages} {379--382} (\bibinfo {year}
  {2019})}\BibitemShut {NoStop}%
\bibitem [{\citenamefont {Wille}\ \emph {et~al.}(2020)\citenamefont {Wille},
  \citenamefont {Jiang}, \citenamefont {Bünermann}, \citenamefont {Wodtke},
  \citenamefont {Behler},\ and\ \citenamefont {Kandratsenka}}]{Wille2020}%
  \BibitemOpen
  \bibfield  {author} {\bibinfo {author} {\bibfnamefont {S.}~\bibnamefont
  {Wille}}, \bibinfo {author} {\bibfnamefont {H.}~\bibnamefont {Jiang}},
  \bibinfo {author} {\bibfnamefont {O.}~\bibnamefont {Bünermann}}, \bibinfo
  {author} {\bibfnamefont {A.~M.}\ \bibnamefont {Wodtke}}, \bibinfo {author}
  {\bibfnamefont {J.}~\bibnamefont {Behler}},\ and\ \bibinfo {author}
  {\bibfnamefont {A.}~\bibnamefont {Kandratsenka}},\ }\bibfield  {title}
  {\enquote {\bibinfo {title} {An experimentally validated neural-network
  potential energy surface for h-atom on free-standing graphene in full
  dimensionality},}\ }\href {https://doi.org/10.1039/D0CP03462B} {\bibfield
  {journal} {\bibinfo  {journal} {Physical Chemistry Chemical Physics}\
  }\textbf {\bibinfo {volume} {22}},\ \bibinfo {pages} {26113--26120} (\bibinfo
  {year} {2020})}\BibitemShut {NoStop}%
\bibitem [{\citenamefont {Behler}\ and\ \citenamefont
  {Parrinello}(2007)}]{Behler2007}%
  \BibitemOpen
  \bibfield  {author} {\bibinfo {author} {\bibfnamefont {J.}~\bibnamefont
  {Behler}}\ and\ \bibinfo {author} {\bibfnamefont {M.}~\bibnamefont
  {Parrinello}},\ }\bibfield  {title} {\enquote {\bibinfo {title} {Generalized
  neural-network representation of high-dimensional potential-energy
  surfaces},}\ }\href {https://doi.org/10.1103/PhysRevLett.98.146401}
  {\bibfield  {journal} {\bibinfo  {journal} {Phys. Rev. Lett.}\ }\textbf
  {\bibinfo {volume} {98}},\ \bibinfo {pages} {146401} (\bibinfo {year}
  {2007})}\BibitemShut {NoStop}%
\bibitem [{\citenamefont {Shi}\ \emph {et~al.}(2023)\citenamefont {Shi},
  \citenamefont {Schr\"oder}, \citenamefont {Meyer}, \citenamefont {Pelaez},
  \citenamefont {Wodtke}, \citenamefont {Golibrzuch}, \citenamefont
  {Sch\"onemann}, \citenamefont {Kandratsenka},\ and\ \citenamefont
  {Gatti}}]{shi23:194102}%
  \BibitemOpen
  \bibfield  {author} {\bibinfo {author} {\bibfnamefont {L.}~\bibnamefont
  {Shi}}, \bibinfo {author} {\bibfnamefont {M.}~\bibnamefont {Schr\"oder}},
  \bibinfo {author} {\bibfnamefont {H.-D.}\ \bibnamefont {Meyer}}, \bibinfo
  {author} {\bibfnamefont {D.}~\bibnamefont {Pelaez}}, \bibinfo {author}
  {\bibfnamefont {A.}~\bibnamefont {Wodtke}}, \bibinfo {author} {\bibfnamefont
  {K.}~\bibnamefont {Golibrzuch}}, \bibinfo {author} {\bibfnamefont {A.-M.}\
  \bibnamefont {Sch\"onemann}}, \bibinfo {author} {\bibfnamefont
  {A.}~\bibnamefont {Kandratsenka}},\ and\ \bibinfo {author} {\bibfnamefont
  {F.}~\bibnamefont {Gatti}},\ }\bibfield  {title} {\enquote {\bibinfo {title}
  {{Quantum and classical molecular dynamics for H atom scattering from
  graphene}},}\ }\href {https://doi.org/10.1063/5.0176655} {\bibfield
  {journal} {\bibinfo  {journal} {J.~Chem.\ Phys.}\ }\textbf {\bibinfo {volume}
  {159}},\ \bibinfo {pages} {194102} (\bibinfo {year} {2023})}\BibitemShut
  {NoStop}%
\bibitem [{\citenamefont {Shi}\ \emph {et~al.}(2024)\citenamefont {Shi},
  \citenamefont {Schr\"oder}, \citenamefont {Meyer}, \citenamefont {Pelaez},
  \citenamefont {Wodtke}, \citenamefont {Golibrzuch}, \citenamefont
  {Sch\"onemann}, \citenamefont {Kandratsenka},\ and\ \citenamefont
  {Gatti}}]{shi23:059901}%
  \BibitemOpen
  \bibfield  {author} {\bibinfo {author} {\bibfnamefont {L.}~\bibnamefont
  {Shi}}, \bibinfo {author} {\bibfnamefont {M.}~\bibnamefont {Schr\"oder}},
  \bibinfo {author} {\bibfnamefont {H.-D.}\ \bibnamefont {Meyer}}, \bibinfo
  {author} {\bibfnamefont {D.}~\bibnamefont {Pelaez}}, \bibinfo {author}
  {\bibfnamefont {A.}~\bibnamefont {Wodtke}}, \bibinfo {author} {\bibfnamefont
  {K.}~\bibnamefont {Golibrzuch}}, \bibinfo {author} {\bibfnamefont {A.-M.}\
  \bibnamefont {Sch\"onemann}}, \bibinfo {author} {\bibfnamefont
  {A.}~\bibnamefont {Kandratsenka}},\ and\ \bibinfo {author} {\bibfnamefont
  {F.}~\bibnamefont {Gatti}},\ }\bibfield  {title} {\enquote {\bibinfo {title}
  {{Erratum: “Quantum and classical molecular dynamics for H atom scattering
  from graphene” [J. Chem. Phys. 159, 194102 (2023)]}},}\ }\href
  {https://doi.org/10.1063/5.0227504} {\bibfield  {journal} {\bibinfo
  {journal} {J.~Chem.\ Phys.}\ }\textbf {\bibinfo {volume} {161}},\ \bibinfo
  {pages} {059901} (\bibinfo {year} {2024})}\BibitemShut {NoStop}%
\bibitem [{\citenamefont {Schröder}(2020)}]{mccpd}%
  \BibitemOpen
  \bibfield  {author} {\bibinfo {author} {\bibfnamefont {M.}~\bibnamefont
  {Schröder}},\ }\bibfield  {title} {\enquote {\bibinfo {title} {Transforming
  high-dimensional potential energy surfaces into a canonical polyadic
  decomposition using monte carlo methods},}\ }\href
  {https://doi.org/10.1063/1.5140085} {\bibfield  {journal} {\bibinfo
  {journal} {Journal of Chemical Physics}\ }\textbf {\bibinfo {volume} {152}}
  (\bibinfo {year} {2020}),\ 10.1063/1.5140085}\BibitemShut {NoStop}%
\bibitem [{\citenamefont {Meyer}, \citenamefont {Manthe},\ and\ \citenamefont
  {Cederbaum}(1990)}]{mey90:73}%
  \BibitemOpen
  \bibfield  {author} {\bibinfo {author} {\bibfnamefont {H.-D.}\ \bibnamefont
  {Meyer}}, \bibinfo {author} {\bibfnamefont {U.}~\bibnamefont {Manthe}},\ and\
  \bibinfo {author} {\bibfnamefont {L.~S.}\ \bibnamefont {Cederbaum}},\
  }\bibfield  {title} {\enquote {\bibinfo {title} {The multi-configurational
  time-dependent {H}artree approach},}\ }\href@noop {} {\bibfield  {journal}
  {\bibinfo  {journal} {Chem. Phys. Lett.}\ }\textbf {\bibinfo {volume}
  {165}},\ \bibinfo {pages} {73--78} (\bibinfo {year} {1990})}\BibitemShut
  {NoStop}%
\bibitem [{\citenamefont {Manthe}, \citenamefont {Meyer},\ and\ \citenamefont
  {Cederbaum}(1992)}]{man92:3199}%
  \BibitemOpen
  \bibfield  {author} {\bibinfo {author} {\bibfnamefont {U.}~\bibnamefont
  {Manthe}}, \bibinfo {author} {\bibfnamefont {H.-D.}\ \bibnamefont {Meyer}},\
  and\ \bibinfo {author} {\bibfnamefont {L.~S.}\ \bibnamefont {Cederbaum}},\
  }\bibfield  {title} {\enquote {\bibinfo {title} {Wave-packet dynamics within
  the multiconfiguration {H}artree framework: General aspects and application
  to {NOCl}},}\ }\href@noop {} {\bibfield  {journal} {\bibinfo  {journal}
  {J.~Chem.\ Phys.}\ }\textbf {\bibinfo {volume} {97}},\ \bibinfo {pages}
  {3199--3213} (\bibinfo {year} {1992})}\BibitemShut {NoStop}%
\bibitem [{\citenamefont {Beck}\ \emph {et~al.}(2000)\citenamefont {Beck},
  \citenamefont {J{\"a}ckle}, \citenamefont {Worth},\ and\ \citenamefont
  {Meyer}}]{bec00:1}%
  \BibitemOpen
  \bibfield  {author} {\bibinfo {author} {\bibfnamefont {M.~H.}\ \bibnamefont
  {Beck}}, \bibinfo {author} {\bibfnamefont {A.}~\bibnamefont {J{\"a}ckle}},
  \bibinfo {author} {\bibfnamefont {G.~A.}\ \bibnamefont {Worth}},\ and\
  \bibinfo {author} {\bibfnamefont {H.-D.}\ \bibnamefont {Meyer}},\ }\bibfield
  {title} {\enquote {\bibinfo {title} {The multiconfiguration time-dependent
  {H}artree method: {A} highly efficient algorithm for propagating
  wavepackets.}}\ }\href@noop {} {\bibfield  {journal} {\bibinfo  {journal}
  {Phys. Rev.}\ }\textbf {\bibinfo {volume} {324}},\ \bibinfo {pages} {1--105}
  (\bibinfo {year} {2000})}\BibitemShut {NoStop}%
\bibitem [{\citenamefont {Meyer}, \citenamefont {Gatti},\ and\ \citenamefont
  {Worth}(2009)}]{mey09:book}%
  \BibitemOpen
  \bibinfo {editor} {\bibfnamefont {H.-D.}\ \bibnamefont {Meyer}}, \bibinfo
  {editor} {\bibfnamefont {F.}~\bibnamefont {Gatti}},\ and\ \bibinfo {editor}
  {\bibfnamefont {G.~A.}\ \bibnamefont {Worth}},\ eds.,\ \href@noop {} {\emph
  {\bibinfo {title} {{Multidimensional Quantum Dynamics: MCTDH Theory and
  Applications}}}}\ (\bibinfo  {publisher} {Wiley-VCH},\ \bibinfo {address}
  {Weinheim},\ \bibinfo {year} {2009})\BibitemShut {NoStop}%
\bibitem [{\citenamefont {Worth}\ \emph {et~al.}()\citenamefont {Worth},
  \citenamefont {Beck}, \citenamefont {J{\"a}ckle}, \citenamefont {Vendrell},\
  and\ \citenamefont {Meyer}}]{mctdh:MLpackage}%
  \BibitemOpen
  \bibfield  {author} {\bibinfo {author} {\bibfnamefont {G.~A.}\ \bibnamefont
  {Worth}}, \bibinfo {author} {\bibfnamefont {M.~H.}\ \bibnamefont {Beck}},
  \bibinfo {author} {\bibfnamefont {A.}~\bibnamefont {J{\"a}ckle}}, \bibinfo
  {author} {\bibfnamefont {O.}~\bibnamefont {Vendrell}},\ and\ \bibinfo
  {author} {\bibfnamefont {H.-D.}\ \bibnamefont {Meyer}},\ }\href@noop {}
  {}\bibinfo {howpublished} {{The MCTDH Package, Version 8.2, (2000). H.-D.
  Meyer, Version 8.3 (2002), Version 8.4 (2007). O. Vendrell and H.-D. Meyer
  {V}ersion 8.5 (2013). H.-D. Meyer, Version 8.6 (2021). Versions 8.5 and 8.6
  contain the ML-MCTDH algorithm. Used versions: 8.6.3 (Jan 2023). {S}ee
  http://mctdh.uni-hd.de/}}\BibitemShut {NoStop}%
\bibitem [{\citenamefont {Auerbach}\ \emph {et~al.}(2020)\citenamefont
  {Auerbach}, \citenamefont {Janke}, \citenamefont {Kammler}, \citenamefont
  {Kandratsenka},\ and\ \citenamefont {Wille}}]{mdt2git}%
  \BibitemOpen
  \bibfield  {author} {\bibinfo {author} {\bibfnamefont {D.~J.}\ \bibnamefont
  {Auerbach}}, \bibinfo {author} {\bibfnamefont {S.~M.}\ \bibnamefont {Janke}},
  \bibinfo {author} {\bibfnamefont {M.}~\bibnamefont {Kammler}}, \bibinfo
  {author} {\bibfnamefont {A.}~\bibnamefont {Kandratsenka}},\ and\ \bibinfo
  {author} {\bibfnamefont {S.}~\bibnamefont {Wille}},\ }\href@noop {} {\enquote
  {\bibinfo {title} {Molecular dynamics tian xia 2 (mdt2); program for
  simulating the scattering of atoms and molecules from a surface (github
  repository). available at https://github.com/akandra/md\_tian2},}\ }
  (\bibinfo {year} {2020})\BibitemShut {NoStop}%
\bibitem [{\citenamefont {Meyer}\ \emph {et~al.}(2006)\citenamefont {Meyer},
  \citenamefont {{Le Qu\'er\'e}}, \citenamefont {L\'eonard},\ and\
  \citenamefont {Gatti}}]{mey06:179}%
  \BibitemOpen
  \bibfield  {author} {\bibinfo {author} {\bibfnamefont {H.-D.}\ \bibnamefont
  {Meyer}}, \bibinfo {author} {\bibfnamefont {F.}~\bibnamefont {{Le
  Qu\'er\'e}}}, \bibinfo {author} {\bibfnamefont {C.}~\bibnamefont
  {L\'eonard}},\ and\ \bibinfo {author} {\bibfnamefont {F.}~\bibnamefont
  {Gatti}},\ }\bibfield  {title} {\enquote {\bibinfo {title} {Calculation and
  selective population of vibrational levels with the {M}ulticonfiguration
  {T}ime-{D}ependent {H}artree ({MCTDH}) algorithm},}\ }\href@noop {}
  {\bibfield  {journal} {\bibinfo  {journal} {Chemical Physics}\ }\textbf
  {\bibinfo {volume} {329}},\ \bibinfo {pages} {179--192} (\bibinfo {year}
  {2006})}\BibitemShut {NoStop}%
\bibitem [{\citenamefont {Riss}\ and\ \citenamefont
  {Meyer}(1993)}]{ris93:4503}%
  \BibitemOpen
  \bibfield  {author} {\bibinfo {author} {\bibfnamefont {U.~V.}\ \bibnamefont
  {Riss}}\ and\ \bibinfo {author} {\bibfnamefont {H.-D.}\ \bibnamefont
  {Meyer}},\ }\bibfield  {title} {\enquote {\bibinfo {title} {Calculation of
  resonance energies and widths using the complex absorbing potential
  method.}}\ }\href@noop {} {\bibfield  {journal} {\bibinfo  {journal} {J.
  Phys. B}\ }\textbf {\bibinfo {volume} {26}},\ \bibinfo {pages} {4503}
  (\bibinfo {year} {1993})}\BibitemShut {NoStop}%
\bibitem [{\citenamefont {Riss}\ and\ \citenamefont
  {Meyer}(1996)}]{ris96:1409}%
  \BibitemOpen
  \bibfield  {author} {\bibinfo {author} {\bibfnamefont {U.~V.}\ \bibnamefont
  {Riss}}\ and\ \bibinfo {author} {\bibfnamefont {H.-D.}\ \bibnamefont
  {Meyer}},\ }\bibfield  {title} {\enquote {\bibinfo {title} {Investigation on
  the reflection and transmission properties of complex absorbing
  potentials.}}\ }\href@noop {} {\bibfield  {journal} {\bibinfo  {journal}
  {J.~Chem.\ Phys.}\ }\textbf {\bibinfo {volume} {105}},\ \bibinfo {pages}
  {1409} (\bibinfo {year} {1996})}\BibitemShut {NoStop}%
\bibitem [{\citenamefont {J{\"a}ckle}\ and\ \citenamefont
  {Meyer}(1995)}]{jae95:5605}%
  \BibitemOpen
  \bibfield  {author} {\bibinfo {author} {\bibfnamefont {A.}~\bibnamefont
  {J{\"a}ckle}}\ and\ \bibinfo {author} {\bibfnamefont {H.-D.}\ \bibnamefont
  {Meyer}},\ }\bibfield  {title} {\enquote {\bibinfo {title} {Reactive
  scattering using the multiconfiguration time-dependent {H}artree
  approximation: {G}eneral aspects and application to the collinear
  {H+H$_2\rightarrow$ H$_2$+H} reaction.}}\ }\href@noop {} {\bibfield
  {journal} {\bibinfo  {journal} {J.~Chem.\ Phys.}\ }\textbf {\bibinfo {volume}
  {102}},\ \bibinfo {pages} {5605} (\bibinfo {year} {1995})}\BibitemShut
  {NoStop}%
\bibitem [{\citenamefont {Comon}, \citenamefont {Luciani},\ and\ \citenamefont
  {de~Almeida}(2009)}]{com09:393}%
  \BibitemOpen
  \bibfield  {author} {\bibinfo {author} {\bibfnamefont {P.}~\bibnamefont
  {Comon}}, \bibinfo {author} {\bibfnamefont {X.}~\bibnamefont {Luciani}},\
  and\ \bibinfo {author} {\bibfnamefont {A.~L.~F.}\ \bibnamefont
  {de~Almeida}},\ }\bibfield  {title} {\enquote {\bibinfo {title} {Tensor
  decompositions, alternating least squares and other tales},}\ }\href@noop {}
  {\bibfield  {journal} {\bibinfo  {journal} {Journal of Chemometrics}\
  }\textbf {\bibinfo {volume} {23}},\ \bibinfo {pages} {393} (\bibinfo {year}
  {2009})}\BibitemShut {NoStop}%
\bibitem [{\citenamefont {{de Lathauwer}}\ and\ \citenamefont
  {Nion}(2008)}]{lat08:1067}%
  \BibitemOpen
  \bibfield  {author} {\bibinfo {author} {\bibfnamefont {L.}~\bibnamefont {{de
  Lathauwer}}}\ and\ \bibinfo {author} {\bibfnamefont {D.}~\bibnamefont
  {Nion}},\ }\bibfield  {title} {\enquote {\bibinfo {title} {{Decompositions of
  a Higher-Order Tensor in Block Terms-Part III: Alternating Least Squares
  Algorithms}},}\ }\href {https://doi.org/doi.org/10.1137/070690730} {\bibfield
   {journal} {\bibinfo  {journal} {SIAM J. Matrix Anal. App.}\ }\textbf
  {\bibinfo {volume} {30}},\ \bibinfo {pages} {1067--1083} (\bibinfo {year}
  {2008})}\BibitemShut {NoStop}%
\bibitem [{\citenamefont {Schröder}\ \emph {et~al.}(2022)\citenamefont
  {Schröder}, \citenamefont {Gatti}, \citenamefont {Lauvergnat}, \citenamefont
  {Meyer},\ and\ \citenamefont {Vendrell}}]{Schroeder2022}%
  \BibitemOpen
  \bibfield  {author} {\bibinfo {author} {\bibfnamefont {M.}~\bibnamefont
  {Schröder}}, \bibinfo {author} {\bibfnamefont {F.}~\bibnamefont {Gatti}},
  \bibinfo {author} {\bibfnamefont {D.}~\bibnamefont {Lauvergnat}}, \bibinfo
  {author} {\bibfnamefont {H.-D.}\ \bibnamefont {Meyer}},\ and\ \bibinfo
  {author} {\bibfnamefont {O.}~\bibnamefont {Vendrell}},\ }\bibfield  {title}
  {\enquote {\bibinfo {title} {The coupling of the hydrated proton to its first
  solvation shell},}\ }\href@noop {} {\bibfield  {journal} {\bibinfo  {journal}
  {Nat. Comm.}\ }\textbf {\bibinfo {volume} {13}},\ \bibinfo {pages} {6170}
  (\bibinfo {year} {2022})}\BibitemShut {NoStop}%
\bibitem [{\citenamefont {Kosztin}, \citenamefont {Faber},\ and\ \citenamefont
  {Schulten}(1996)}]{Kosztin1996}%
  \BibitemOpen
  \bibfield  {author} {\bibinfo {author} {\bibfnamefont {I.}~\bibnamefont
  {Kosztin}}, \bibinfo {author} {\bibfnamefont {B.}~\bibnamefont {Faber}},\
  and\ \bibinfo {author} {\bibfnamefont {K.}~\bibnamefont {Schulten}},\
  }\bibfield  {title} {\enquote {\bibinfo {title} {Introduction to the
  diffusion monte carlo method},}\ }\href {https://doi.org/10.1119/1.18168}
  {\bibfield  {journal} {\bibinfo  {journal} {American Journal of Physics}\
  }\textbf {\bibinfo {volume} {64}},\ \bibinfo {pages} {633--644} (\bibinfo
  {year} {1996})}\BibitemShut {NoStop}%
\bibitem [{\citenamefont {Wang}\ and\ \citenamefont {Thoss}(2003)}]{Wang2003}%
  \BibitemOpen
  \bibfield  {author} {\bibinfo {author} {\bibfnamefont {H.}~\bibnamefont
  {Wang}}\ and\ \bibinfo {author} {\bibfnamefont {M.}~\bibnamefont {Thoss}},\
  }\bibfield  {title} {\enquote {\bibinfo {title} {Multilayer formulation of
  the multiconfiguration time-dependent hartree theory},}\ }\href
  {https://doi.org/10.1063/1.1580111} {\bibfield  {journal} {\bibinfo
  {journal} {Journal of Chemical Physics}\ }\textbf {\bibinfo {volume} {119}},\
  \bibinfo {pages} {1289--1299} (\bibinfo {year} {2003})}\BibitemShut {NoStop}%
\bibitem [{\citenamefont {Manthe}(2008)}]{man08:164116}%
  \BibitemOpen
  \bibfield  {author} {\bibinfo {author} {\bibfnamefont {U.}~\bibnamefont
  {Manthe}},\ }\bibfield  {title} {\enquote {\bibinfo {title} {A multilayer
  multiconfigurational time-dependent {H}artree approach for quantum dynamics
  on general potential energy surfaces},}\ }\href@noop {} {\bibfield  {journal}
  {\bibinfo  {journal} {J.~Chem.\ Phys.}\ }\textbf {\bibinfo {volume} {128}},\
  \bibinfo {pages} {164116} (\bibinfo {year} {2008})}\BibitemShut {NoStop}%
\bibitem [{\citenamefont {Vendrell}\ and\ \citenamefont
  {Meyer}(2011)}]{Vendrell2011}%
  \BibitemOpen
  \bibfield  {author} {\bibinfo {author} {\bibfnamefont {O.}~\bibnamefont
  {Vendrell}}\ and\ \bibinfo {author} {\bibfnamefont {H.-D.}\ \bibnamefont
  {Meyer}},\ }\bibfield  {title} {\enquote {\bibinfo {title} {Multilayer
  multiconfiguration time-dependent hartree method: Implementation and
  applications to a henon-heiles hamiltonian and to pyrazine},}\ }\href
  {https://doi.org/10.1063/1.3535541} {\bibfield  {journal} {\bibinfo
  {journal} {Journal of Chemical Physics}\ }\textbf {\bibinfo {volume} {134}}
  (\bibinfo {year} {2011}),\ 10.1063/1.3535541}\BibitemShut {NoStop}%
\bibitem [{\citenamefont {Mendive-Tapia}, \citenamefont {Meyer},\ and\
  \citenamefont {Vendrell}(2023)}]{men23:1144}%
  \BibitemOpen
  \bibfield  {author} {\bibinfo {author} {\bibfnamefont {D.}~\bibnamefont
  {Mendive-Tapia}}, \bibinfo {author} {\bibfnamefont {H.-D.}\ \bibnamefont
  {Meyer}},\ and\ \bibinfo {author} {\bibfnamefont {O.}~\bibnamefont
  {Vendrell}},\ }\bibfield  {title} {\enquote {\bibinfo {title} {{Optimal mode
  combinations in the Multiconfiguration Time-Dependent Hartree method through
  multivariate statistics: Factor analysis and hierarchical clustering}},}\
  }\href {https://doi.org/10.1021/acs.jctc.2c01089} {\bibfield  {journal}
  {\bibinfo  {journal} {J.~Chem.\ Theory Comput.}\ }\textbf {\bibinfo {volume}
  {19}},\ \bibinfo {pages} {1144} (\bibinfo {year} {2023})}\BibitemShut
  {NoStop}%
\bibitem [{\citenamefont {Golibrzuch}\ \emph {et~al.}(2022)\citenamefont
  {Golibrzuch}, \citenamefont {Walpole}, \citenamefont {Schönemann},\ and\
  \citenamefont {Wodtke}}]{Exp2022}%
  \BibitemOpen
  \bibfield  {author} {\bibinfo {author} {\bibfnamefont {K.}~\bibnamefont
  {Golibrzuch}}, \bibinfo {author} {\bibfnamefont {V.}~\bibnamefont {Walpole}},
  \bibinfo {author} {\bibfnamefont {A.-M.}\ \bibnamefont {Schönemann}},\ and\
  \bibinfo {author} {\bibfnamefont {A.~M.}\ \bibnamefont {Wodtke}},\ }\bibfield
   {title} {\enquote {\bibinfo {title} {Generation of sub-nanosecond h atom
  pulses for scattering from single-crystal epitaxial graphene},}\ }\href
  {https://doi.org/10.1021/acs.jpca.2c05364} {\bibfield  {journal} {\bibinfo
  {journal} {The Journal of Physical Chemistry A}\ }\textbf {\bibinfo {volume}
  {126}},\ \bibinfo {pages} {8101--8110} (\bibinfo {year} {2022})}\BibitemShut
  {NoStop}%
\bibitem [{\citenamefont {Bonfanti}\ \emph {et~al.}(2015)\citenamefont
  {Bonfanti}, \citenamefont {Jackson}, \citenamefont {Hughes}, \citenamefont
  {Burghardt},\ and\ \citenamefont {Martinazzo}}]{Bonfanti2015}%
  \BibitemOpen
  \bibfield  {author} {\bibinfo {author} {\bibfnamefont {M.}~\bibnamefont
  {Bonfanti}}, \bibinfo {author} {\bibfnamefont {B.}~\bibnamefont {Jackson}},
  \bibinfo {author} {\bibfnamefont {K.~H.}\ \bibnamefont {Hughes}}, \bibinfo
  {author} {\bibfnamefont {I.}~\bibnamefont {Burghardt}},\ and\ \bibinfo
  {author} {\bibfnamefont {R.}~\bibnamefont {Martinazzo}},\ }\bibfield  {title}
  {\enquote {\bibinfo {title} {Quantum dynamics of hydrogen atoms on graphene.
  ii. sticking},}\ }\href@noop {} {\bibfield  {journal} {\bibinfo  {journal}
  {The Journal of Chemical Physics}\ }\textbf {\bibinfo {volume} {143}},\
  \bibinfo {pages} {124704} (\bibinfo {year} {2015})}\BibitemShut {NoStop}%
\bibitem [{\citenamefont {Bonfanti}, \citenamefont {Achilli},\ and\
  \citenamefont {Martinazzo}(2018)}]{Bonfanti_2018}%
  \BibitemOpen
  \bibfield  {author} {\bibinfo {author} {\bibfnamefont {M.}~\bibnamefont
  {Bonfanti}}, \bibinfo {author} {\bibfnamefont {S.}~\bibnamefont {Achilli}},\
  and\ \bibinfo {author} {\bibfnamefont {R.}~\bibnamefont {Martinazzo}},\
  }\bibfield  {title} {\enquote {\bibinfo {title} {Sticking of atomic hydrogen
  on graphene},}\ }\href {https://doi.org/10.1088/1361-648X/aac89f} {\bibfield
  {journal} {\bibinfo  {journal} {Journal of Physics: Condensed Matter}\
  }\textbf {\bibinfo {volume} {30}},\ \bibinfo {pages} {283002} (\bibinfo
  {year} {2018})}\BibitemShut {NoStop}%
\end{thebibliography}%

%

\end{document}